\documentclass{aa}  

\usepackage{graphicx}
\usepackage{txfonts}
\usepackage{natbib}
\bibpunct{(}{)}{;}{a}{}{,} 

\usepackage{url}

\begin{document}

\title{The halo of M49 and its environment as traced by planetary nebulae populations\thanks{Based on data collected at Subaru Telescope, which is operated by the National Astronomical Observatory of Japan under programme S14A-006.}}

\author{J. Hartke\inst{1} \and
        M. Arnaboldi\inst{1,2} \and
        A. Longobardi\inst{3,4} \and
        O. Gerhard\inst{4} \and
        K. C. Freeman\inst{5} \and
        S. Okamura\inst{6} \and
        F. Nakata\inst{7}
        }

\institute{European Southern Observatory,
        Karl-Schwarzschild-Str. 2, 85748 Garching, Germany\\
        \email{[jhartke;marnabol]@eso.org}
        \and
        INAF, Observatory of Pino Torinese, Turin, Italy 
        \and
        Kavli Institute for Astronomy and Astrophysics,
        Peking University, 5 Yiheyuan Road, Haidian District, Beijing 100871, P. R. China\\
        \email{alongobardi@pku.edu.cn}
        \and
        Max-Planck-Institut für Extraterrestrische Physik, 
        Giessenbachstrasse, 85748 Garching, Germany\\
        \email{gerhard@mpe.mpg.de}
        \and
        RSAA, Mt. Stromlo Observatory, 2611 Canberra, Australia\\
        \email{Kenneth.Freeman@anu.edu.au}
        \and
        Department of Advanced Sciences, Faculty of Science and Engineering, 
        Hosei University, 184-8584 Tokyo, Japan\\
        \email{sadanori.okamura@hosei.ac.jp}
        \and
        Subaru Telescope, National Astronomical Observatory of Japan, 650 N Aohoku Place Hilo HI96720, U.S.A.\\
        \email{nakata@naoj.org}
}
\date{\today}

\abstract{The galaxy M49 (NGC 4472) is the brightest early-type galaxy in the Virgo Cluster. It is located in Subcluster B and has an unusually blue, metal-poor outer halo. Planetary nebulae (PNe) are excellent tracers of diffuse galaxy and intragroup light.} 
{We aim to  present a photometric survey of  PNe in the galaxy's extended halo to characterise its PN population, as well as the surrounding intragroup light (IGL) of the Subcluster B.}
{PNe were identified based on their bright [OIII]5007 \AA\ emission and absence of a broad-band continuum through automated detection techniques.}
{We identify $738$ PNe out to a radius of $\sim155\;\mathrm{kpc}$ from M49's centre from which we define a complete sample of $624$ PNe within a limiting magnitude of $m_\mathrm{5007,lim}=28.8$. Comparing the PN number density to the broad-band stellar surface brightness profile, we find a variation of the PN-specific frequency ($\alpha$-parameter) with radius. The outer halo beyond $60\;\mathrm{kpc}$ has a $3.2$ times higher $\alpha$-parameter compared to the main galaxy halo ($\alpha_{2.5,\mathrm{inner}}^{\mathrm{M49}}=(3.20\pm 0.43)\times 10^{-9}\,\mathrm{PN}\,L^{-1}_{\odot,\mathrm{bol}}$), which is likely due to contribution from the surrounding blue IGL. We use the Planetary Nebulae Luminosity Function (PNLF) as an indicator of  distance and stellar population. Its slope, which correlates empirically with galaxy type, varies within the inner halo. In the eastern quadrant of M49, the PNLF slope is shallower, indicating an additional localised, bright PN population following an accretion event, likely that of the dwarf irregular galaxy VCC1249. We also determined a distance modulus of $\mu_{\mathrm{PNLF}}=31.29^{+0.07}_{-0.08}$ for M49, corresponding to a physical distance of $18.1\pm 0.6\,\mathrm{Mpc}$, which agrees with a recent surface-brightness fluctuations distance.}
{The PN populations in the outer halo of M49 are consistent with the presence of a main S\'ersic galaxy halo with a slight $(B-V)$ colour gradient of $10^{-4}\;\mathrm{mag}\;\mathrm{arcsec}^{-1}$ surrounded by intragroup light with a very blue colour of $(B-V)=0.25$ and a constant surface brightness $\mu_V=28.0\;\mathrm{mag\;arcsec}^{-2}$.}


\maketitle
%

\section{Introduction}
\defcitealias{2013A&A...558A..42L}{L13}
\defcitealias{2009ApJS..182..216K}{K09}
\defcitealias{2010ApJ...715..972J}{J10}
\defcitealias{2015A&A...581A..10C}{C15}
In the context of hierarchical structure formation, galaxies grow by mergers and accretion of smaller structures \citep[e.g.][]{1991ApJ...379...52W,2002NewA....7..155S,2012ApJ...744...63O} that leave long-lasting signatures in the galaxies' outermost regions, because of long dynamical timescales (of the order of a Gyr at 50 kpc). Early-type galaxies (ETGs) predominantly consist of a very old population of stars that are believed to have formed on a short timescale \citep[$\sim 1$ Gyr,][and references therein]{2005MNRAS.360.1355T}. The strong size evolution of massive ETGs, that is, of a factor of between two and four from $z\sim2$ \citep[e.g.][]{2005ApJ...626..680D, 2008ApJ...677L...5V} can be explained with a two-phase formation scenario for which the initial phase of violent star formation is followed by accretion events \citep{2010ApJ...725.2312O, 2015ApJ...799..184P}, whose stellar populations are also old. Hence the distribution of stars and their dynamics at very large radii around massive ETGs should contain important evidence of their accretion histories \citep{2013MNRAS.434.3348C, 2016ApJ...833..158C}.

In nearby galaxies, the halos can be mapped with absorption-line spectroscopy of red giant branch (RGB) stars, providing spatial and line-of-sight (LOS) velocity distributions \citep[e.g.][]{2009Natur.461...66M, 2013MNRAS.432..832C, 2016ApJ...823...19C}. While RGB stars are an excellent tracer to map the halos of the Milky Way (MW), M31, or Centaurus A, it becomes too costly to carry out absorption-line spectroscopy for faint sources with apparent magnitudes $m_V > 23.5\,\mathrm{mag}$. 

Integral-field spectroscopy has played a pivotal role in characterising the two-dimensional kinematics of ETGs \citep{2016ARA&A..54..597C} within the high surface brightness regions ($1-2 r_{\mathrm{e}}$). To reach larger distances, discrete tracers are needed to study the halos of bright ETGs.  Two common discrete tracers at distances larger than $2r_{\mathrm{e}}$ are planetary nebulae (PNe) and globular clusters (GCs). The latter can be separated into red (metal-rich) and blue (metal-poor) populations, which possess different spatial, velocity, and angular momentum distributions \citep{2013MNRAS.436.1322C, 2013MNRAS.428..389P, 2014ApJ...796...62H}. The GC systems in nearby bright ETGs have been studied in multiple surveys (e.g. SLUGSS: \citet{2012MNRAS.426.1475U}, NGVS: \citet{2014ApJ...794..103D}).

PNe are a population of dying asymptotic giant branch (AGB) stars in the zero-age main sequence mass range of $1-8\,\mathrm{M}_{\odot}$. The core star strongly emits in the UV. This radiation ionises the envelope of gas surrounding the core star, which re-emits 10\% of the UV radiation in one optical line: the [OIII]5007\AA\ line. It can still be detected in distant galaxies and serves as an identifier of PNe \citep{1992ApJ...389...27D}. Observational evidence suggests that PNe trace stellar light \citep[e.g.][]{1989ApJ...339...53C,2009MNRAS.394.1249C,2013MNRAS.432.1010C} and that their angular momentum distribution is that of the stars \citep[e.g.][]{1995ApJ...449..592H, 1996ApJ...472..145A, 1998ApJ...507..759A, 2001ApJ...563..135M}. As their luminosity-specific frequency does not vary much with galaxy type \citep{2006MNRAS.368..877B}, they are an effective tracer of substructures, which have not been detected in previous imaging or integrated absorption-line spectroscopy,  as for example the stellar stream in the M31 disc \citep{2006MNRAS.369..120M} and the crown in M87 \citep{2015A&A...579L...3L}.

In order to relate the observed properties of the PN population in external galaxies with the properties of the stellar population (e.g. age and metallicity), one characterises the PN population by the luminosity-specific PN number ($\alpha$-parameter) and the shape of the PN luminosity function (PNLF). Observational evidence suggests that the $\alpha$-parameter, which relates the stellar luminosity with the number of PN, is further correlated with the integrated $(B - V)$ colour of the galaxy \citep{2006MNRAS.368..877B}. \citet{1989ApJ...339...53C} analytically describe the shape of the PNLF based on the LF of PNe in M31. The shape of the PNLF is governed by a bright cut-off at $m^{\star}$. The absolute magnitude $M^{\star}$ of this cut-off is invariant with galaxy type and can therefore be used as a secondary distance indicator to the respective galaxy \citep{2002ApJ...577...31C}. 

The Virgo Cluster is an interesting target as there are many indicators for a complex dynamical history. It is one of the closest cluster environments and has already been targeted by a number of PN surveys \citep{1998ApJ...492...62C, 1998ApJ...503..109F, 2004ApJ...615..196F, 2002AJ....123..760A, 2003AJ....125..514A,  2005AJ....129.2585A, 2009A&A...507..621C}.
It is composed of multiple subgroups (Virgo A, B, and C), that were first identified by \citet{1987AJ.....94..251B, 1993A&AS...98..275B}, centred on the ETGs M87, M49, and M60. The X-ray emission of the hot and dense intergalactic medium is highest at the centres of the subclusters, where the concentration of quiescent galaxies peaks \citep{1999A&A...343..420S}. \citet{2014A&A...570A..69B} observe a colour-segregation effect in the (sub)cluster: red galaxies are located in the high-density regions, whereas blue star-forming galaxies are found in the outskirts of the cluster. Galaxies at the centre of subcluster A are devoid of dusty and gaseous phases of the ISM, while subcluster B mainly consists of star-forming galaxies.

Deep wide-field imaging reveals intricate structures of diffuse intracluster light (ICL) surrounding the bright ETGs, suggesting a hierarchical build-up \citep{2005ApJ...631L..41M, 2017ApJ...834...16M}. This is also reflected in the distribution of intracluster PNe, which show field-to-field variations \citep{2004ApJ...615..196F, 2005AJ....129.2585A, 2009A&A...507..621C}. The colours of ICL features around the cluster core near the cD galaxy NGC 4486 (M87) are similar to those of M87's halo itself, suggesting that they may have formed from similar progenitors \citep{2010ApJ...720..569R}. 

The galaxy M87 has been the target of a successful pilot survey combining narrow-band photometry from Suprime-Cam at the Subaru telescope \citep[][hereafter \citetalias{2013A&A...558A..42L}]{2013A&A...558A..42L} with high-resolution spectroscopy with FLAMES at the VLT \citep{2015A&A...579A.135L}, uncovering the transition from an inner PN-scarce (old and metal-rich) galaxy to an outer PN-rich, accretion-driven component (young and metal-poor). \citet{2015A&A...579L...3L} identified a recent satellite accretion event through a chevron-like PN structure, that also manifests itself as excess light in the M87 outer halo.

We extended this survey to NGC 4472 (M49), the brightest elliptical galaxy in the Virgo Cluster, residing at the centre of subcluster B. The X-ray bow shock to the north of the galaxy suggests that it is falling into the centre of the Virgo Cluster \citep{1996ApJ...471..683I}. M49 possesses a system of shells along the major axis, suggestive of past accretion events of satellite galaxies \citep{2010ApJ...715..972J, 2015A&A...581A..10C}. Further substructure is associated with the dwarf galaxy VCC 1249 (UGC 7636). Optical and infrared observations combined with H$\alpha$ measurements show signs of a recent interaction between the dwarf and M49 \citep{2012A&A...543A.112A}. \citet{2003ApJ...591..850C} mapped the dynamics of M49's GC system within $400^{\prime\prime}$ ($33\;\mathrm{kpc}$), and a more extended photometric GC study mapped M49's halo in its entirety, finding a smooth GC density distribution with radius \citep{2014ApJ...794..103D}.

This paper is organised as follows: in Sect. \ref{sec:data} we describe the photometric PN survey and the data reduction procedure. The selection and validation of PN candidates is described in Sect. \ref{sec:selection}. We then study the relation between the azimuthally averaged spatial distribution of the PN candidates and the stellar surface brightness of M49 in Sect. \ref{sec:density}. In Sect. \ref{sec:PNLF} we present the PNLF of M49 and the corresponding inferred distance. In Sect. \ref{sec:halo_variation} we trace variation of the PNLF in M49's halo with our PN catalogue and discuss whether these are signatures of substructure in Sect. \ref{sec:discussion}. We summarise and conclude in Sect. \ref{sec:summary}. If not stated otherwise, we assume a distance of 17.1 Mpc \citep{2007ApJ...655..144M} to M49 that corresponds to a physical scale of $82\;\mathrm{pc}$ per $1''$. 

\section{The Suprime-Cam M49 PN survey}
\label{sec:data}
This survey expands the ongoing effort to study PNe as tracers in cluster and group environments. We use the same techniques as \citetalias{2013A&A...558A..42L}, who observed M87 at the centre of Virgo's subcluster A using the same instrumental configuration. In order to detect PN candidates, we used a narrow-band filter centred on the [OIII]5007\AA-line and measure the colour excess of spatially unresolved sources with respect to the continuum, measured in the broad $V$-band.

\subsection{Imaging and observations}
The observations were carried out with the Suprime-Cam at the prime focus of the 8.2 m Subaru telescope \citep{2002PASJ...54..833M} in May 2014. The 10k $\times$ 8k mosaic camera covers an area of $34^{\prime}\times27^{\prime}$, corresponding to a scale of $0.2^{\prime\prime}$ per pixel. The halo of M49 can thus be observed out to a radial distance of $155$ kpc with a single pointing. 

As we identified PNe using the on-off-band technique (see Sect. \ref{sec:selection} for further details), we observe M49 through a narrow-band [OIII] filter ($\lambda_c = 5029\,\AA$, $\Delta \lambda = 74\,\AA$, on-band) and a broad-band V-filter ($\lambda_c = 5500\,\AA$, $\Delta \lambda = 956\,\AA$, off-band). The on-band image consists of 13 dithered exposures with a total exposure time of three hours and the off-band image of 14 dithered exposures with a total exposure time of 1.4 hours. The exposure time was chosen such that PNe with an apparent narrow-band magnitude of 2.5 mag from the bright cut-off $m^{\star}_{5007} = 26.79$ can still be detected assuming a distance modulus of 30.8 for Virgo.
The observations were carried out in one night, with photometric conditions. Throughout the whole night the seeing was better than $0''.7$ and the airmass of the observations varied between 1.1 and 1.5. 

\subsection{Data reduction, astrometry and flux calibration}
\label{ssec:data_red}
The data were reduced using the instrument pipeline SDFRED\footnote{\url{http://www.naoj.org/Observing/Instruments/SCam/sdfred}} for Subaru's Suprime-Cam. As this data set will be used to provide positions of PN candidates for a spectroscopic follow-up, accurate astrometry is needed. We improved the accuracy of the images using the IRAF\footnote{IRAF is distributed by the National Optical Astronomy Observatories, which are operated by the Association of Universities for Research in Astronomy, Inc., under cooperative agreement with the National Science Foundation} package \texttt{imcoords} with the 2MASS catalogue as an astrometric reference frame. 

In order to flux calibrate the broad-band and narrow-band frames to the $AB$ magnitude system, we observed standard stars through the same filters used for the survey, using the spectrophotometric HILT600 and BD332642 stars for the [OIII]-band and Landolt stars for the $V$-band. The zero points for the [OIII] and $V$-band frames normalised to a 1 second exposure are $Z_{\mathrm{[OIII]}} = 24.51 \pm 0.04$ and $Z_{\mathrm{V}} = 27.48 \pm 0.01$. These values are consistent with those obtained by \citetalias{2013A&A...558A..42L}, and \citet{2009A&A...507..621C}, who use the same filters for the Suprime Cam narrow-band photometry.

The integrated flux $F_{5007}$ from the [OIII] line can be related to the magnitude $m_{5007}$ using the following relation \citep{1989ApJ...339...39J}
\begin{equation}
        m_{5007} = -2.5\log_{10} F_{5007} - 13.74,
\end{equation}
where the flux is in units of $\mathrm{ergs}\,\mathrm{cm}^{-2}\mathrm{s}^{-1}$.
\citet{2003AJ....125..514A} determine the relation between the $AB$ and $m_{5007}$ magnitudes to be 
\begin{equation}
        m_{5007} = m_{AB} + 2.49
\end{equation}
for the narrow-band filter in use.

The final images are the astrometrically and photometrically calibrated, stacked images in the [OIII] and $V$-band. We fitted the psf using the IRAF task \texttt{psf}. The best-fit profile is a Gaussian analytical function with a radius $r = 2.36$ pixels and a standard deviation $\sigma = 2.35$ pixels. We also computed the PSF fit in four different image regions separately showing that the PSF does not vary across the image.

\section{Selection of the PN candidates and catalogue extraction}
\label{sec:selection}
PNe have a bright [OIII]5007\AA\ and no continuum emission.\footnote{If a faint continuum emission is detected at the position of [OIII] emission, it is likely to be caused by background residuals or crowding.}
Therefore extragalactic PNe can be identified as objects detected in the on-band [OIII] image, which are not detected in the off-band continuum images, or in other words, which have an excess in [OIII] - V colour. Furthermore PNe are unresolved at extragalactic distances, therefore we only considered point-like objects for analysis. We used the automatic selection procedure developed and validated in \citet{2002AJ....123..760A, 2003AJ....125..514A}, subsequently applied to studies of PN populations in Virgo's diffuse ICL \citep[e.g.][]{2009A&A...507..621C} and in Virgo ellipticals. The selection procedure has been tailored to Suprime-Cam data by \citetalias{2013A&A...558A..42L}. As the selection procedure is vital to the properties of the extracted catalogue we review the main steps in the following section and refer to Appendix \ref{app:selection} for further information regarding the basic catalogue extraction.

\subsection{Catalogue pre-processing}

Throughout the following analysis, we used SExtractor \citep{1996A&AS..117..393B} as a tool to detect and carry out photometry of the sources on the images. The narrow- and broad-band magnitudes $m_\mathrm{n}$ and $m_\mathrm{b}$ are measured in dual-image mode for sources detected on the narrow-band image.
In order to determine the \textbf{limiting magnitude} of our survey, we populated the images with a synthetic PN population that follows a Ciardullo-like PNLF and determine the recovery fraction as a function of magnitude. The magnitude at which the recovery fraction drops below $50\%$ is adopted as the limiting magnitude;
on the narrow-band image it is $m_{\mathrm{n,lim,cat}} = 26.8$ ($AB$ magnitude) or $m_{\mathrm{5007,lim,cat}} = 29.3$ (see Sect. \ref{app:extraction}).

There are different \textbf{image artefacts} that can lead to false detections of PNe. We therefore masked the pixel columns affected by dithering as well as regions with a high background value (e.g. due to charge transfer or saturated stars). Due to the high contribution of diffuse galaxy light we furthermore excluded objects detected within a major-axis radius of $r_{\mathrm{major}} = 159^{\prime\prime}$ (corresponding to $13$ kpc) from the galaxy's centre.

\subsection{Object selection}
\begin{figure}
\begin{center}
        \includegraphics[width = 8.8cm]{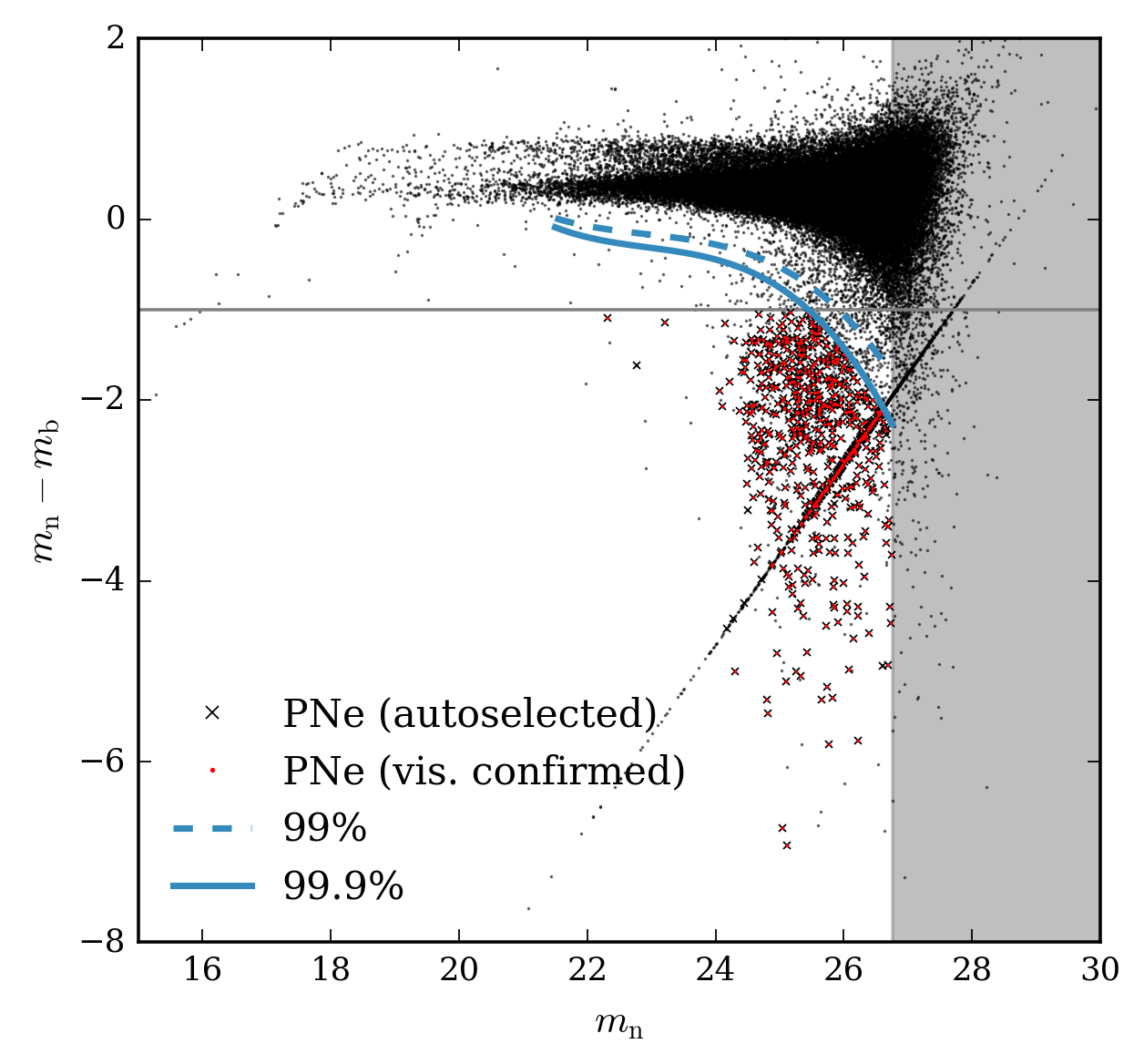}
        \caption{CMD for all sources in the M49 Suprime Cam field. Objects below the horizontal line are emission line objects with an $EW_{\mathrm{obs}} \leq 110\;\AA$ and colour excess. The solid (dashed) blue lines indicate the regions above which $99.9\%$ ($99\%$) of simulated continuum objects fall. Objects fainter than the limiting magnitude fall into the grey region. The candidate PNe from the automatic selection are denoted by crosses and those confirmed by visual inspection of the images by red dots.}
        \label{fig:cmdsel}
\end{center}
\end{figure}

\begin{figure}
\begin{center}
        \includegraphics[width = 8.8cm]{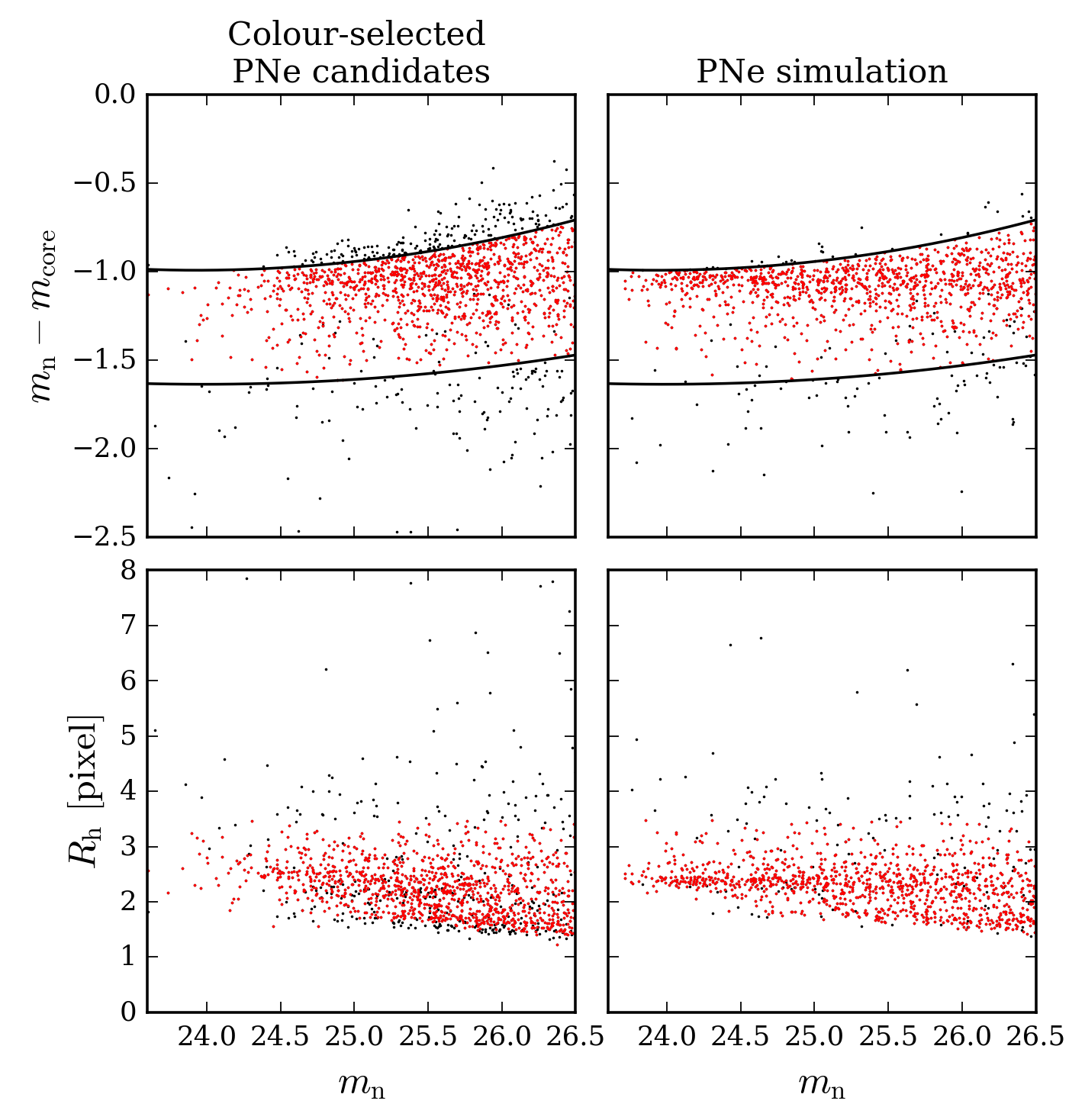}
        \caption{Point-source test: The left column shows the colour-selected PNe candidates and the right the simulated PN-population. The top row shows $m_{\mathrm{n}} - m_{\mathrm{core}}$ vs. $m_{\mathrm{n}}$, the black lines indicate the region in which 95\% of the simulated population fall. The bottom row shows $R_{\mathrm{h}}$ as a function of $m_{\mathrm{n}}$. We impose a limit of 3.5 pixel on $R_\mathrm{h}$, such that 95\% of the simulated population is included. Objects which satisfy both criteria are plotted in red and are termed point-like.}
        \label{fig:pointsource}
\end{center}
\end{figure}
We selected objects as PN candidates based on their position on the colour-magnitude diagram (CMD) and their spatially unresolved light distribution. The CMD selection criteria are shown in Fig. \ref{fig:cmdsel}. We selected objects with a colour excess of $m_{\mathrm{n}} - m_{\mathrm{b}} \leq -1$ that are brighter than the survey's limiting magnitude  $m_{\mathrm{n,lim,cat}} = 26.8$. In order to limit the contamination by foreground stars we then constrained our sample to those objects that fall below the $99.9\%$ confidence limits of a simulated continuum population.

In Fig. \ref{fig:cmdsel} and Fig. \ref{fig:pointsource} we show the CMD-selected sources that are classified as unresolved or point-like sources. Based on the light distribution of the simulated population, point-sources satisfy the following two criteria: these sources (i) have a half-light radius of $1 < R_{\mathrm{h}} < 3.5$ pixel, where the upper limit corresponds to the $95\%$-percentile of the simulated population, and
(ii) they fall in the region where the difference between $m_{\mathrm{n}}$ and $m_{\mathrm{core}}$ is within the $95\%$-limits of the simulated population (see Sect. \ref{app:pointlike}).

\subsection{Catalogue completeness}
\label{ssec:phot_compl}
Evaluating the completeness of our catalogue is a critical step to be carried out before we can use it for further investigation (see Sect. \ref{app:completeness} for further details). We again made use of our synthetic PN population and determined its recovery as a function of radius to assess its spatial completeness ($c_\mathrm{spatial}(r)$) and as a function of magnitude for the photometric completeness. For the latter, we independently determined the detection and colour completeness ($c_\mathrm{detection}(m_{5007})$ and $c_\mathrm{colour}(m_{5007})$ respectively). 

\subsection{Catalogue validation and visual inspection}
\label{ssec:vis_insp}
\begin{figure}
\begin{center}
        \includegraphics[width = 8.8cm]{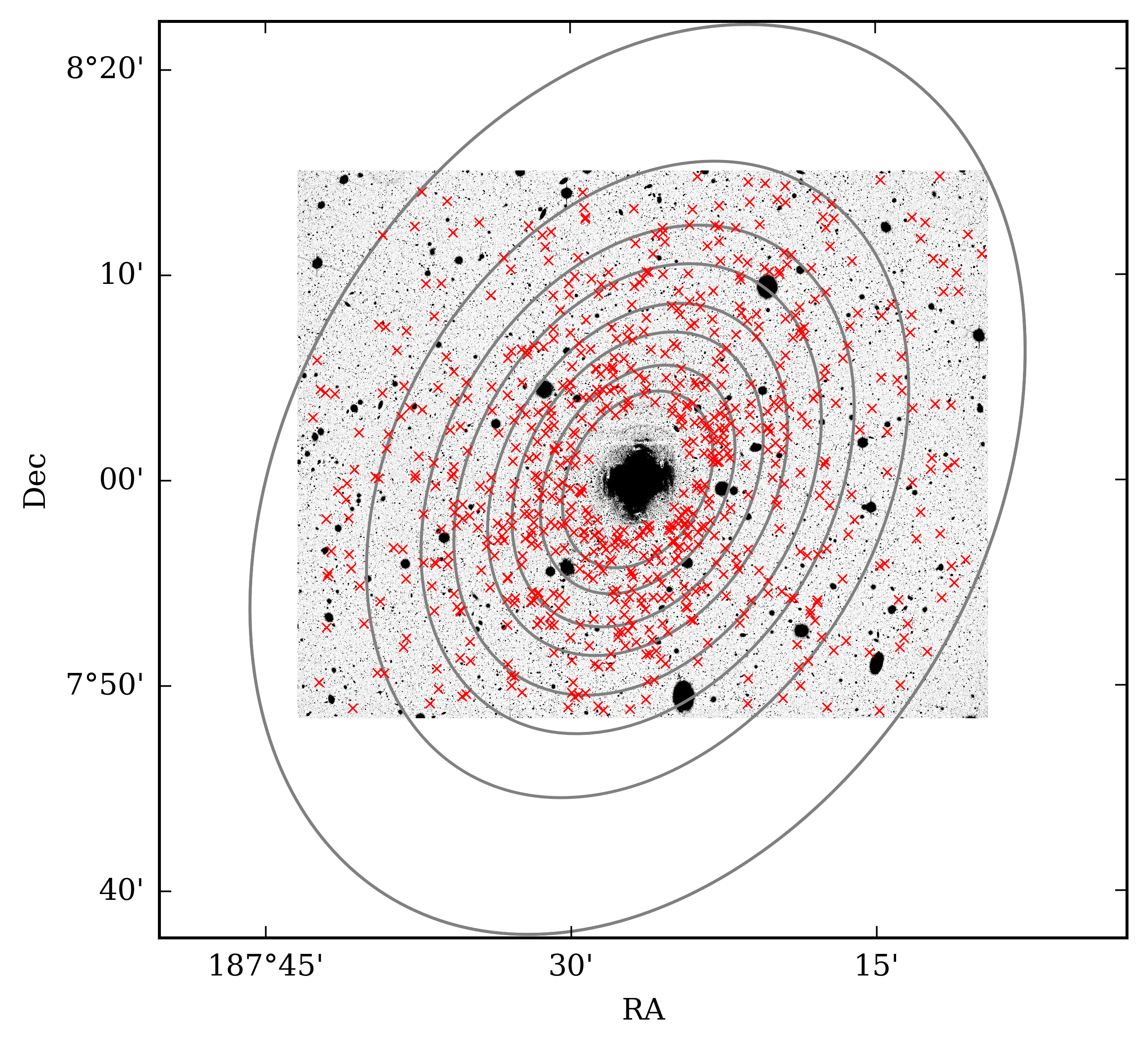}
        \caption{[OIII] image of M49 with PN candidates overplotted in red. The elliptical bins have $PA = - 31^{\circ}$ and a constant ellipticity $\epsilon = 0.28$ \citepalias{2009ApJS..182..216K} and were defined such that they contain the same number of objects. North is up, east is to the left.}
        \label{fig:bins}
\end{center}
\end{figure}

\begin{figure}
\begin{center}
        \includegraphics[width = 8.8cm]{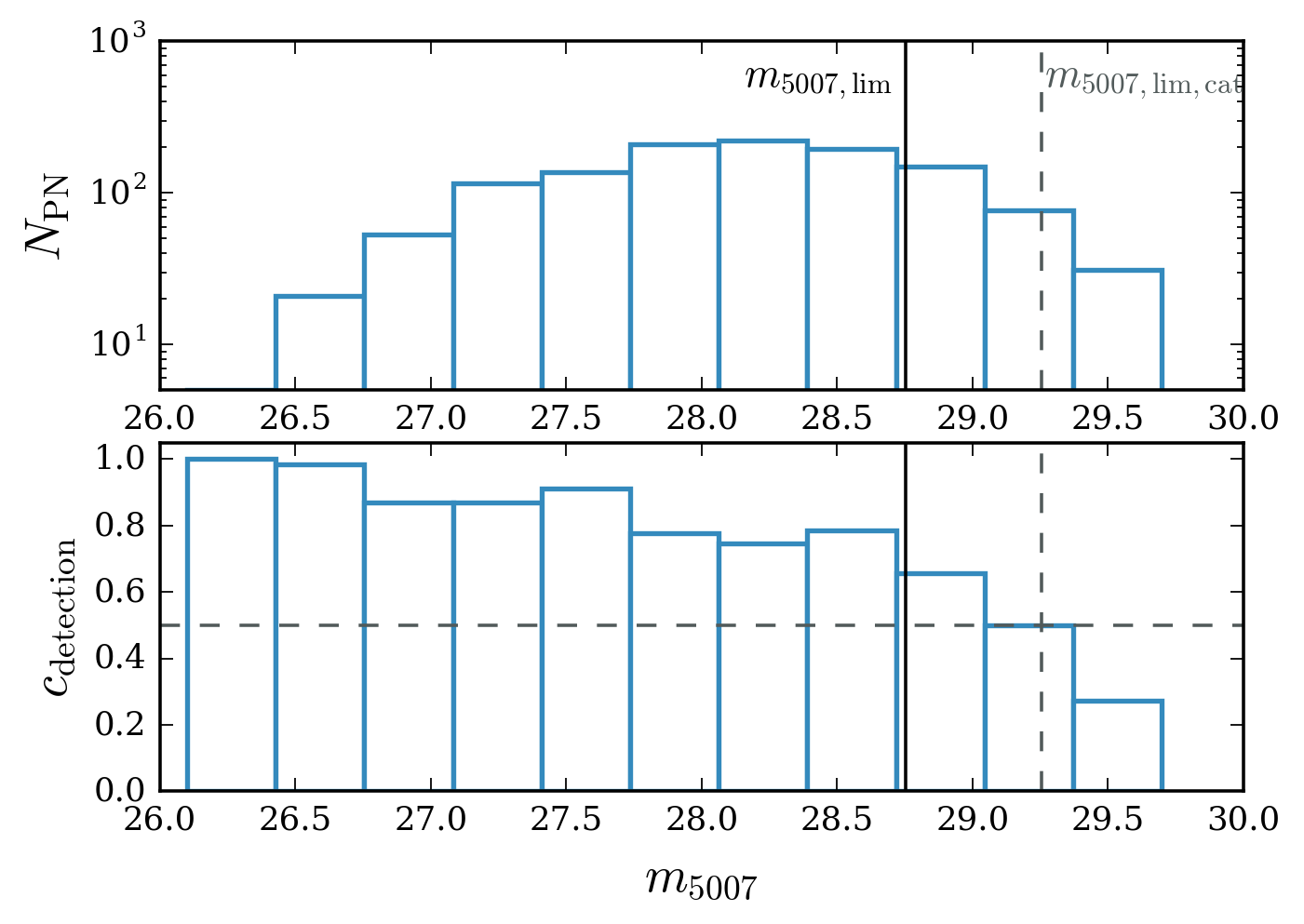}
        \caption{Observed number counts of PN as function of magnitude (top panel) and detection completeness of a simulated PN population (bottom panel). The dashed horizontal line denotes the $50\%$ recovery limit that results in the limiting magnitude $m_\mathrm{5007,lim,cat} = 29.3$ indicated by the dashed vertical line. The solid vertical line denotes the conservative limiting magnitude $m_{\mathrm{5007,lim}} = 28.8$ based on the decrease of the observed PN-number counts.}
        \label{fig:completeness}
\end{center}
\end{figure}

The photometric catalogue obtained from the automatic selection was visually inspected and any remaining spurious detections were removed. Spurious detections amounted to $\sim3\%$ of the automatically extracted sources, the final catalogue contains 735 sources. The visually confirmed PN candidates are shown in Fig. \ref{fig:cmdsel}. The spatial distribution of the extracted and visually confirmed objects is shown in Fig. \ref{fig:bins}, superposed on the continuum-subtracted mosaic image.

While the limiting magnitude based on completeness estimations is $m_{\mathrm{5007,lim,cat}} = 29.3$ (see Sect. \ref{ssec:phot_compl}), we introduce a more conservative limit of $m_{\mathrm{5007,lim}} = 28.8$ based on the morphology of M49's PNLF (see Sect. \ref{sec:PNLF}). At fainter magnitudes, the PN-number counts decline strongly (Fig. \ref{fig:completeness}). Given that the number of PNe increases exponentially as a function of magnitude beyond the cut-off in old stellar populations \citep{1989ApJ...339...53C, 2015A&A...579L...3L}, such behaviour is indicative of photometric incompleteness. Our final complete sample within  $m_{\mathrm{5007,lim}} = 28.8$ consists of 624 PNe.

\subsection{Possible sources of contamination}

\subsubsection{Contamination by faint continuum objects}
As we describe in Sect. \ref{ssec:colour_sel}, we limited our selection of objects to those that fall below the $99.9\%$-line of a simulated continuum population in order to minimise the contamination by foreground objects, for example, faint Milky Way stars. However, some of these objects will be scattered below this line. We estimated the fraction of foreground contaminants in the colour-selected sample by determining the total number of observed foreground stars down to $m_\mathrm{n,lim}$ and assuming that $0.1\%$ of these are scattered into the sample of colour-selected PNe. This results in $47$ objects, which corresponds to a contribution of $8\%$ of the extracted sample of 624 PNe. 

\subsubsection{Contamination by background galaxies}
\defcitealias{2007ApJ...667...79G}{G07}
\label{ssec:lyalpha}
Another source of contamination are faint background Ly-$\alpha$ galaxies, that emit at the same wavelength as PNe in the Virgo Cluster, if at redshift $z = 3.1$. In order to quantify this effect, we use the Ly-$\alpha$ luminosity function (LF) determined by \citet[][hereafter \citetalias{2007ApJ...667...79G}]{2007ApJ...667...79G}, who carried out a deep survey for $z = 3.1$ Ly-$\alpha$ emission-line galaxies. The LF is characterised by a Schechter-function \citep{1976ApJ...203..297S}. More recent surveys \citep[e.g.][]{2012ApJ...744..110C} agree with this LF within $0.1$ mag. For a detailed description, we refer the reader to \citetalias{2013A&A...558A..42L} who estimated the number of background contaminants for two Suprime-Cam fields covering Virgo's central galaxy M87. An important validation of this approach is the confirmation of the number of estimated contaminants during the spectroscopic follow-up of PN candidates \citep{2015A&A...579A.135L}.
As our survey is half a magnitude deeper than that of \citetalias{2007ApJ...667...79G}, we extrapolate the LF to a limiting magnitude of $m_{\mathrm{5007,lim}} = 28.8$. We expect 310 Ly-$\alpha$ emitters, which corresponds to  $(29.8\pm6.0)\%$ of the completeness-corrected sample. The corresponding LF is shown in Fig. \ref{fig:PNLF_nocorr}. 

Another source of background contaminants are [OII]3727\AA\ emitters at redshift $z = 0.345$. The colour-selection criterion $m_{\mathrm{n}} - m_{\mathrm{b}} < -1$ corresponds to an observed equivalent width of $EW_{\mathrm{obs}} > 110\,\AA$, which reduces the contribution from [OII] emitters, as no [OII] emitters beyond $EW_{\mathrm{obs}} > 95\,\AA$ have been observed \citep{1990MNRAS.244..408C, 1997ApJ...481...49H, 1998ApJ...504..622H}. \citetalias{2007ApJ...667...79G} employ a similar colour-selection criterion ($\Delta m_{\mathrm{G07}} \lesssim -1$) and consider the fraction of [OII] contaminants at redshift $z = 0.345$ negligible. Therefore the contribution of background [OII] emitters with $EW_{\mathrm{obs}} > 110\,\AA$ is accounted for when using the \citetalias{2007ApJ...667...79G} Ly-$\alpha$ LF.

\subsection{Comparison with other PN-surveys of M49}
\begin{figure}
\begin{center}
        \includegraphics[width = 8.8cm]{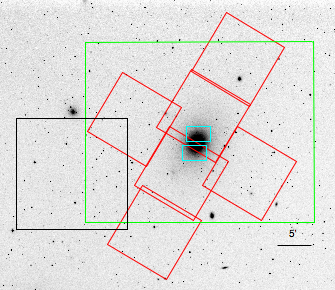}
        \caption{DSS image of M49 with PN survey fields overplotted. Green: this survey, cyan: \citet{1990ApJ...356..332J}, black: \citet{2003ApJS..145...65F}, red: PN.S \citep{PNS_Pulsoni}}
        \label{fig:M49_fields}
\end{center}
\end{figure}

\begin{figure}
\begin{center}
        \includegraphics[width = 8.8cm]{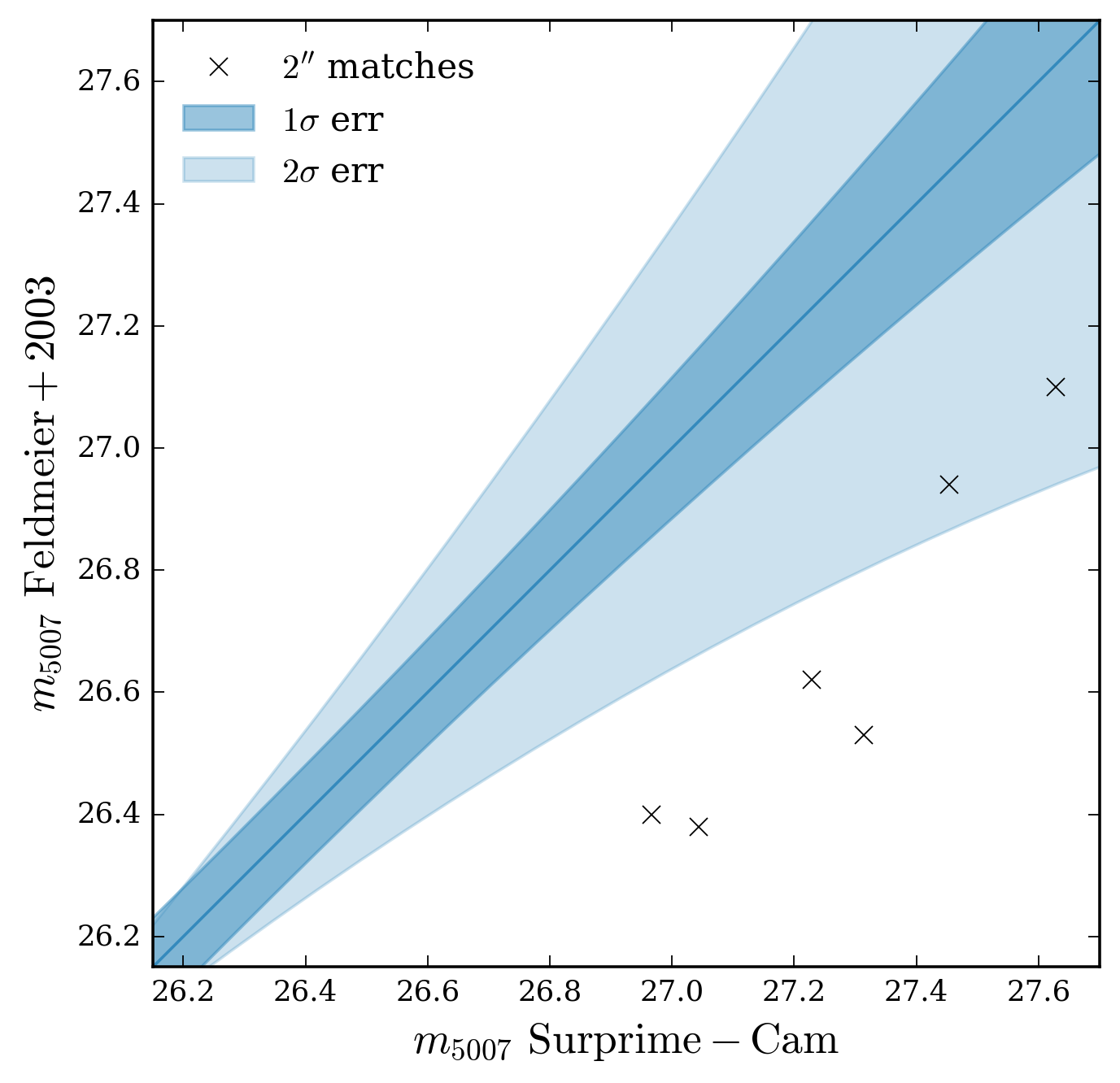}
        \caption{Magnitude comparison and photometric error of the matched emission-line objects between our catalogue and \citet{2003ApJS..145...65F}.}
        \label{fig:feldmeier_comp}
\end{center}
\end{figure}
The first survey of PNe in M49 was carried out by \citet{1990ApJ...356..332J}. They detected 54 objects with [OIII] excess, of which 27 belong to the statistically complete sample as defined by the authors. This survey consists of two $3.'5\times2.'2$ CFHT fields that cover the central region of M49 and are indicated in Fig. \ref{fig:M49_fields} (cyan rectangles). As the majority of this region is masked out in our deep images, only one common PN candidate is identified. We will discuss the comparison between our PNLF  and that of \citet{1990ApJ...356..332J} in Sect. \ref{sec:PNLF}.

Our survey also overlaps with a previous photometric study of intracluster PNe \citep{2003ApJS..145...65F, 2004ApJ...615..196F}. The field that overlaps with our survey is denoted by a black rectangle in Fig. \ref{fig:M49_fields} and has been observed with the Prime Focus CCD camera at the Kitt Peak National Observatory (KPNO) 4m telescope. The survey depth in the [OIII] is $m_{5007} = 28$ mag. A cross-matching of the survey with our catalogue identifies six common objects. The comparison of the [OIII] magnitudes is shown in Fig. \ref{fig:feldmeier_comp}. An average zero-point offset of 0.5 mag is found. A previous comparison of M87's PN sample from from Suprime Cam photometry \citetalias{2013A&A...558A..42L}  with PNe identified by \citet{2003ApJS..145...65F} found an offset in magnitude zero-point that equalled to 0.3 mag. 

Currently there is no published spectroscopic PNe data for M49 available. However, we  have access to unpublished data from the Planetary Nebula Spectrograph \citep[PN.S,][]{2002PASP..114.1234D} as part of a survey of ETGs \citep[][in prep.]{PNS_Pulsoni}. Of the six PN.S fields ($10^{\prime}\times10^{\prime}$, indicated in red in Fig. \ref{fig:M49_fields}) four are aligned along the major axis and two along the minor axis, covering $18^{\prime}$ and $15^{\prime}$ along the respective axis, extending a little further along the major axis than our survey. In total, $465$ PNe have been detected, of which $226$ fall into the unmasked regions of our observations. The PN.S is capable of detecting PNe in the Virgo Cluster down to $\Delta m = 1.5$ from the bright cut-off. We therefore matched only PNe in this magnitude range with the PN.S catalogue. There are $207$ PNe in common within a matching radius of $5^{\prime\prime}$, which serves as a first spectroscopic confirmation of these sources. 

\section{The radial PN number density profile and the PN luminosity-specific frequency}
\label{sec:density}
In the following section we determine the radial PN number density profile and compare it to stellar surface brightness profiles derived from broad-band photometric studies of the M49 halo. Empirically, it has been shown that the PN number density profile follows the stellar light in elliptical \citep[L13;][]{2009MNRAS.394.1249C,  2015A&A...579L...3L} and S0 galaxies \citep{2013A&A...549A.115C, 2013MNRAS.432.1010C}. In the last decade, there have been a number of deep photometric surveys of the Virgo cluster. In this section, we will discuss the surface brightness profile and the 2D properties of the stellar light in M49's extended halo. In a cluster environment, surface brightness and colour profiles do not only provide important evidence about the galaxy halo, but also about its environment, the ICL. Our goal is to link the structural components in the light, which are signalled by a change of slope in the radial gradients of the SB profile, isophotal twists, and different colour gradients, with variations in the PN population parameters as the number density profile, PNLF slopes and the $\alpha$-parameter.
\subsection{Stellar surface photometry of M49}
\label{ssec:photometry}

The stellar surface brightness profile of ETGs is well-described by a S\'ersic function \citep{1963BAAA....6...41S} of the form 
\begin{equation}
        I(r) = 10^{-0.4\mu_e} 10^{-b ((r/r_e)^{1/n}-1)},
\end{equation}
with $b = 0.868n_V - 0.142$.
For M49, \citet[][hereafter \citetalias{2009ApJS..182..216K}]{2009ApJS..182..216K} determine the major-axis fit parameters in the $V$-band to be $\mu_{e,V} = 23.371^{+0.171}_{-0.144}\;\mathrm{mag}\,\mathrm{arcsec}^{-2}$, $r_{e,V} = 269.291^{+23.634}_{-18.570}\,\mathrm{arcsec}$, and $n_V = 5.992^{+0.314}_{-0.292}$. The S\'ersic profile is also fit to data from a deep $V$-band imaging survey of diffuse light in the Virgo Cluster \citet{2010ApJ...715..972J}, the fit-parameters agreeing within $1-2 \sigma$ with \citetalias{2009ApJS..182..216K}. The effective radius derived from integration of the 2D profile is $r_e = 194.44\pm17.0^{\prime\prime}$.

More extended photometry from the VEGAS survey \citep[][hereafter \citetalias{2015A&A...581A..10C}]{2015A&A...581A..10C} is also well-fit by a S\'ersic profile with an effective radius of $r_e = 152^{\prime\prime} \pm 7^{\prime\prime}$ in both the $g$- and $i$-bands.\footnote{In order to match the $g$-band photometry to the $V$-band photometry of \citetalias{2009ApJS..182..216K}, we adopted a constant offset of $-0.35$ as determined by \citetalias{2015A&A...581A..10C}.} 
The $g$-band profile changes slope beyond a major-axis distance of $r \sim 915^{\prime\prime}$. A hint of this flattening at large radii can also be seen in the $i$-band and was also already observed in the surface brightness profile derived from photographic plates by \citet{1994A&AS..106..199C}. Figure \ref{eqn:density} shows the $V$- and $g$-band stellar surface brightness profiles of \citetalias{2009ApJS..182..216K} and \citetalias{2015A&A...581A..10C} as well as the S\'ersic fit determined by \citetalias{2009ApJS..182..216K} . 

Beyond a major-axis distance of $r \sim 250^{\prime\prime}$ the orientation and morphology of the isophotes change. The position angle, which is at constant $-20^{\circ}$ for the main galaxy halo, rapidly decreases from $-20^{\circ}$ to $-40^{\circ}$ in the $g$- and $V$-band and rapidly increases from $-20^{\circ}$ to $-5^{\circ}$ in the $i$-band. Furthermore, the isophotal ellipticity $\epsilon$ changes: the $g$- and $V$-band isophotes become more elongated, reaching a maximum ellipticity of $\epsilon \approx 0.3$, while the $i$-band isophotes become nearly round \citepalias[$\epsilon \approx 0$,][see Fig. 7 therein]{2015A&A...581A..10C}. 

\citet{2005ApJ...618..195G} suggest that the rapid variation of the isophotes' parameters as well as the flattening of the surface brightness profile at large radii is likely due to a population of ICL. The flattening of the surface brightness profile occurs at a surface brightness level of $\mu_g \sim 27\,\mathrm{mag}\,\mathrm{arcsec}^{-2}$ \citepalias{2015A&A...581A..10C}, which is compatible with the surface brightness where a ICL-induced change of slope is observed in a sample of stacked SDSS galaxy clusters between redshift $z=0.2$ and $0.3$. \citep{2005MNRAS.358..949Z}. In the case of M49, which is located at the heart of the Virgo Subcluster B, this extra light is probably not intracluster light from the main Virgo Cluster, but rather intragroup light \citep[IGL,][]{2006ApJ...648..936R, 2005ApJ...618..195G} of the subcluster B.

The colour profile of M49 has a clear gradient that is shallow in the red, inner regions and steepens towards the bluer outskirts \citep[][C15]{2013ApJ...764L..20M}. The transition from shallow to steep gradient occurs at $r \sim 200^{\prime\prime}$, similar to where the isophotes start to change ellipticity and orientation. There are at least two causes for such a colour gradient, which might be interlinked. It can be interpreted as a metallicity gradient from a metal-rich to a metal-poor component, but also as an age gradient from an older to a younger population. This supports the assumption of a two-component galaxy halo with a change of halo composition in the outer part.
When comparing the results of the VEGAS survey \citepalias{2015A&A...581A..10C} with the ones of \citet{2013ApJ...764L..20M, 2017ApJ...834...16M}, the VST broadband images show an extended outer envelope that is responsible for the change of slope in the surface brightness profile beyond $r^{1/4} = 5.5$. The most reliable measurement of the envelope's colour is that of \citet{2013ApJ...764L..20M} of $(B - V) = 0.66 \pm 0.02$ at $r^{1/4} = 6.0$.   

In the remainder of this section, we wish to investigate whether we observe similar trends in our PN sample. Signatures are expected to be found in the radial PN number density profile and the $\alpha$-parameters. 

\subsection{The radial PN number density profile}
\begin{figure}
\begin{center}
        \includegraphics[width = 8.8cm]{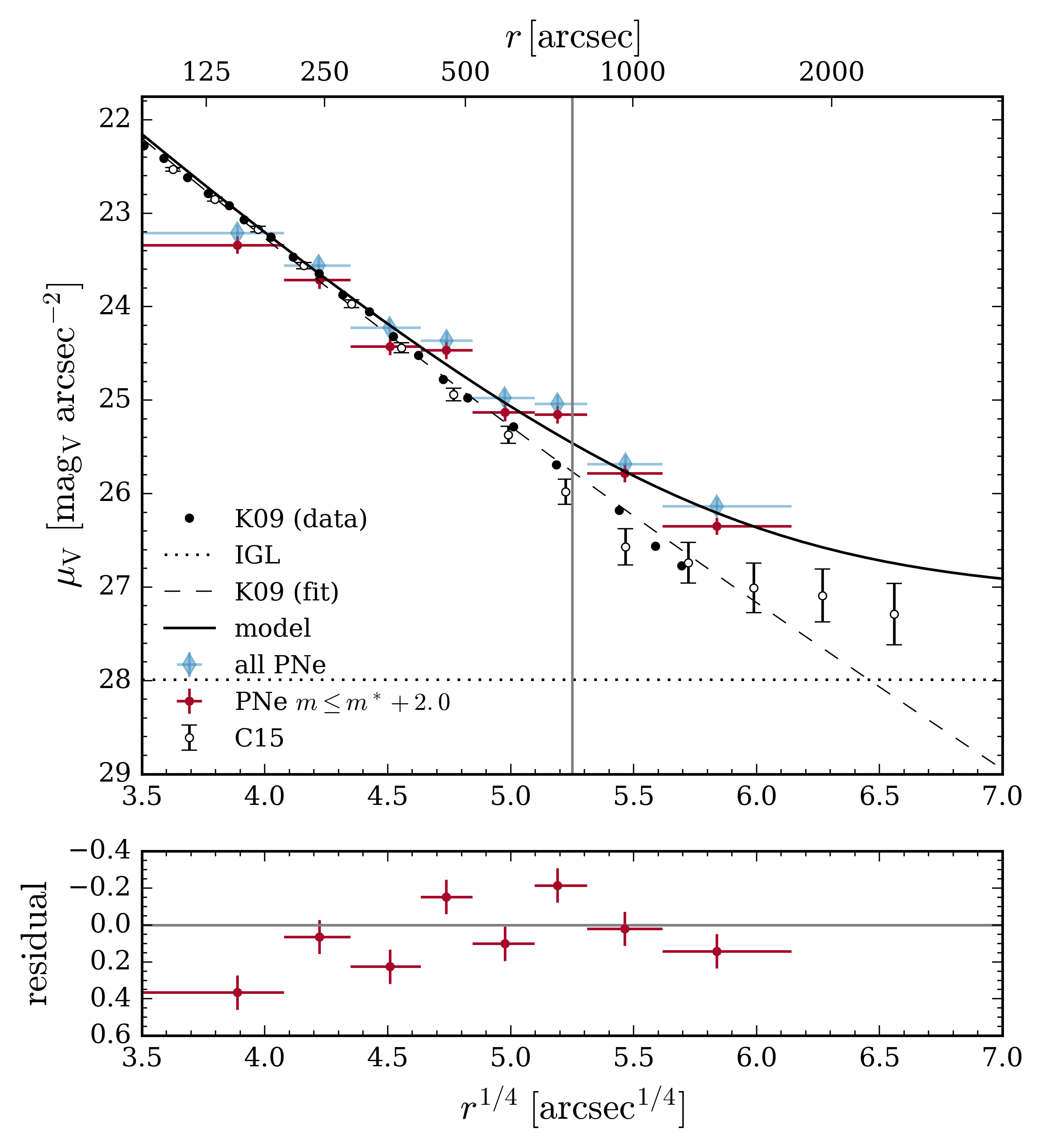}
        \caption{Top panel shows the radial surface brightness profiles in M49: the filled circles denote the surface brightness profile observed by \citetalias{2009ApJS..182..216K} and the open circles are from \citetalias{2015A&A...581A..10C}. The S\'ersic fit \citepalias{2009ApJS..182..216K} is indicated by the dashed line. The dotted horizontal line shows the expected contribution of the IGL. The PN number density profile is denoted with blue diamonds and red circles, where the latter sample has a magnitude limit of $m^{\star} + 2.0$ mag. The two-component model fit to the PN number density profile is shown with the solid black line. The vertical line denotes the transition radius from the inner PN-scarce to the outer PN-rich halo based on the change in the PN number density slope. The bottom panel shows the residuals of the fit.}
        \label{fig:density}
\end{center}
\end{figure}

Figure \ref{fig:bins} shows the distribution of PNe on the sky together with elliptical bins, aligned with the photometric major axis with a $PA = -31^{\circ}$ and a constant ellipticity $\epsilon = 0.28$ as measured from isophotes \citepalias{2009ApJS..182..216K}. The binning has been chosen such that every bin contains the same number of objects ($\sim 70$). The PN logarithmic number density profile is defined as 
\begin{equation}
        \mu_{\mathrm{PN}}(r) = -2.5\log_{10}\Sigma_{\mathrm{PN}}(r) + \mu_0,
        \label{eqn:density}
\end{equation}
where $\mu_0$ is a constant in order to match the PN number density profile with the stellar surface brightness profile and $\Sigma_{\mathrm{PN}}$ is the completeness-corrected PN number density. As detailed in Sect. \ref{ssec:phot_compl} we have to account for colour and spatial incompleteness ($c_{\mathrm{colour}}$ and $c_{\mathrm{spatial}}$ respectively) that have been determined in Sect. \ref{ssec:phot_compl} and detailed values are given in Appendix \ref{app:selection}, Table \ref{tab:photometric} and Table \ref{tab:spatial}. The total number of expected PNe is thus
\begin{equation}
        N_{\mathrm{PN,corr}}(r) = \frac{N_{\mathrm{PN,obs}}(r)}{c_{\mathrm{spatial}}(r)c_{\mathrm{colour}}}
,\end{equation}
and the number density is
\begin{equation}
        \Sigma_{\mathrm{PN}}(r) = \frac{N_{\mathrm{PN,corr}}(r)}{A(r)}.
\end{equation}
The area $A(r)$ is the area of intersection of the elliptical annulus with the field of view of the observation, accounting for the regions that have been masked, which was computed using Monte-Carlo integration techniques.

As explained in Sect. \ref{ssec:lyalpha}, we also need to account for background galaxies at $z = 3.1$ emitting Ly-$\alpha$ radiation redshifted to our narrow-band filter range. We therefore statistically subtracted the contribution of Ly-$\alpha$ emitters from the completeness corrected number of PNe $N_{\mathrm{PN,corr}}$. We assume a homogeneous distribution in the surveyed area, accounting for a $20\%$ variation in the Ly-$\alpha$ LF due to large-scale cosmic variance (cf. App. \ref{sec:lya}). We calculated the number density profile adopting a limiting magnitude of $m_{\mathrm{5007,lim}} = 28.8$ (see Sect. \ref{ssec:vis_insp}). Given that the PNLF bright cut-off is at $m_{5007}^* = 26.8$ (see Sect. \ref{sec:PNLF}), we span a magnitude range of $\Delta m = 2$.

The resulting density profile matched to the V-band surface brightness profile \citepalias{2009ApJS..182..216K} is shown in Fig. \ref{fig:density}. We also computed the number density profile for all PNe down to the automatically determined limiting magnitude $m_{\mathrm{5007,lim}} = 29.3$  and only find a constant offset to the magnitude-limited sample which does not depend on radius.  We cover a radial range of $250^{\prime\prime}$ to $1785^{\prime\prime}$, corresponding to $20$ to $155$ kpc, spanning $9r_e$.\footnote{assuming the effective radius from the 2D profile integration of \citetalias{2009ApJS..182..216K}} 
The error bars account for the counting statistics and the uncertainty on the Ly-$\alpha$ LF. 

The PN number density profile follows the stellar light closely out to a radius of $r \sim 730^{\prime\prime}$. At larger radii it starts to flatten and diverges from the S\'ersic profile. We note that the flattening occurs much earlier than for the stellar surface brightness profile, but seems to have a steeper slope compared to the flattened part of the stellar profile. As we statistically subtracted the contribution from Ly-$\alpha$ background galaxies, we do not expect the flattening to be due to Ly-$\alpha$ emitters \citepalias[see also][Sect. 4.1]{2013A&A...558A..42L}. The origin of the flattening of the logarithmic PN number density is investigated in the following.

\subsection{The PN luminosity-specific number -- the $\alpha$-parameter}
Before we fit a model to the PN number density profile, we need to establish the connection between the PN population and the underlying stellar population. 
The total number of PNe can be directly related with the total bolometric luminosity $L_{\mathrm{bol}}$ of the parent stellar population via the so-called $\alpha$-parameter:
\begin{equation}
        N_{\mathrm{PN}} = \alpha L_{\mathrm{bol}}
        \label{eqn:alphadef}
\end{equation}

As our survey is magnitude-limited to $\Delta m = 2$ magnitudes from the bright cut-off, we first determined the $\alpha$-parameter for that magnitude range, which is defined to be
\begin{equation}
        N_{\mathrm{PN,\Delta m}}=\int_{m^*}^{m^*+\Delta m}\! \mathrm{N}(m) \ \mathrm{d}m = \alpha_{\Delta m}L_{\mathrm{bol}},
        \label{eqn:alphalim}
\end{equation} 
where $N(m)$ is the PNLF and $m^{\star}_{5007} = 26.8$ its apparent bright cut-off magnitude.

If we sample the entire stellar population, we can determine the total number of PNe associated to a parent stellar population with total bolometric luminosity $L_{\mathrm{bol}}$ under the assumption that the luminosity-specific stellar death rate is approximately independent of its parent population's age, metallicity, and initial mass function \citep{1986ASSL..122..195R}:
\begin{equation}
        N_{\mathrm{PN}} = BL_{\mathrm{bol}}\tau_{\mathrm{PN}},
        \label{eqn:PNlifetime}
\end{equation}
where $\tau_{\mathrm{PN}}$ is the PN visibility lifetime and $B$ is the specific evolutionary flux.\footnote{The evolutionary flux quantifies the stellar death rate based on the flow of stars through post-main sequence evolutionary stages. Normalising this quantity by the unit light of the stellar population one gets the specific evolutionary flux \citep[see e.g.][]{2011spug.book.....G}.} Combining Eqs. \eqref{eqn:alphadef} and \eqref{eqn:PNlifetime} one can determine the luminosity-specific PN number:
\begin{equation}
        \alpha = \frac{N_{\mathrm{PN}}}{L_{\mathrm{bol}}} = B\tau_{\mathrm{PN}}
\end{equation}
Simulations of simple stellar populations predict that the variations of $B$ with metallicity or Initial Mass Function (IMF) are small -- values of $B$ are in the range of $1-2 \times 10^{-11}\;\mathrm{L}^{-1}_{\odot}\;\mathrm{yr}^{-1}$ for simulated simple stellar populations \citep{1986ASSL..122..195R} -- thus one can understand the observed values of the $\alpha$-parameter as a proxy for the visibility lifetime $\tau_{\mathrm{PN}}$  of the PNe associated with the parent stellar populations. 
For PN-surveys that do not have the sensitivity to span the entire PNLF, the luminosity specific alpha-parameter provides an estimate of the PN visibility lifetime within the limiting magnitude of the survey.

\subsection{A two-component photometric model for M49}
\label{ssec:2comp}
The flattening of PN number density profile at large radii might be due to an additional outer component. This can be a secondary component of M49's halo, or the diffuse light component of the Virgo Subcluster B (IGL), at whose centre M49 resides \citep{1987AJ.....94..251B, 1993A&AS...98..275B}. We therefore constructed a two-component photometric model in order to reproduce the flattening at large radii. A change of slope is also observed in the stellar surface brightness profile, but at larger radii \citepalias{2015A&A...581A..10C}.

The inner component is dominated by a S\'ersic profile. As the slope of the outer stellar surface brightness profile has not been fitted to date, we followed the suggestion of \citetalias{2013A&A...558A..42L} to use a constant surface brightness for the IGL component. We determined this component's value by determining the differential surface brightness between the last measured value of the flattened stellar profile \citepalias{2015A&A...581A..10C} and the extrapolated S\'ersic fit at that radius and find $\mu_{\mathrm{outer}} = 28.0\,\mathrm{mag}\,\mathrm{arcsec}^{-2}$.

The photometric model for the predicted PN surface density is then
\begin{align}
        \tilde{\Sigma}(r) &= (\alpha_{\Delta m, \mathrm{inner}} I_{\mathrm{inner,bol}}(r) + \alpha_{\Delta m, \mathrm{outer}} I_{\mathrm{outer,bol}}(r))s^2 \\
        &= \alpha_{\Delta m, \mathrm{inner}} \left(I_{\mathrm{inner,bol}}(r) + \left(\frac{\alpha_{\Delta m, \mathrm{outer}}}{\alpha_{\mathrm{\Delta m, inner}}} - 1 \right) I_{\mathrm{\Delta m, outer,bol}}(r)\right)s^2 .
\end{align}
The two components $I_{\mathrm{inner,bol}}(r)$ and $I_{\mathrm{outer,bol}}(r)$ are the bolometric surface brightness profiles related by the $\alpha$-parameters of the respective populations and $s = D/206265$ is a scale factor related to the galaxy's distance $D$. The bolometric correction is done via
\begin{equation}
        I = 10^{-0.4(\mathrm{BC}_V - \mathrm{BC}_{\odot})}10^{-0.4(\mu - K)},
\end{equation}
with the solar bolometric correction $\mathrm{BC}_{\odot} = -0.07$, and $K = 26.4\,\mathrm{mag}\,\mathrm{arcsec}^{-2}$ is the $V$-band conversion factor to physical units $L_{\odot}\,\mathrm{pc}^{-2}$. One can assume a fixed value of $\mathrm{BC}_V = -0.85$ with 10\% accuracy based on the study of stellar population models for different galaxy types \citep[irregular to elliptical, see][]{2006MNRAS.368..877B}.

In order to determine the $\alpha$-parameter of the inner component, we matched PN surface density      (Eq. \eqref{eqn:density}) to the stellar surface brightness profile in the $V$-band \citepalias{2009ApJS..182..216K} in the inner region, where the two profiles follow each other closely. The $\alpha$-parameter can then be determined from the offset $\mu_0$ between the two profiles and applying the bolometric correction:
\begin{equation}
        \alpha_{\mathrm{inner}} = \frac{10^{0.4(\mu_0 - K - (\mathrm{BC}_{\odot}- \mathrm{BC}_V ))}}{s^2}
\end{equation}
Within the chosen magnitude range ($\Delta m = 2$), we determined an offset of $\mu_0 = 15.0 \pm 0.1\,\mathrm{mag}\;\mathrm{arcsec}^{-2}$, corresponding to $\alpha_{2,\mathrm{inner}} = (1.99\pm 0.27) \times 10^{-9}\,\mathrm{PN}\,L^{-1}_{\odot,\mathrm{bol}}$.

We then fitted the two-component model to the observed PN number density profile over the whole radial range. As we have already determined $\alpha_{2,\mathrm{inner}}$, the only free parameter is the ratio of the two parameters $\alpha_{\mathrm{\Delta m, outer}}/\alpha_{\mathrm{\Delta m, inner}}$, which is determined using the fit algorithm \texttt{lmfit} \citep{newville_2014_11813}. 
The best-fit ratio is $\alpha_{\mathrm{2, outer}}/\alpha_{\mathrm{2, inner}} = 3.21 \pm 0.54$ with $\chi^2_{\mathrm{red}} = 2.9$ and the resulting best-fit model is denoted by the black line in Fig. \ref{fig:density}.

A common magnitude range in which $\alpha$-parameters are evaluated is within $\Delta m = 2.5$ magnitudes from the bright cut-off. We can extrapolate $\alpha_{2.5}$ to from a sample that is only complete to a magnitude $m_c < m^{\star} + 2.5$ using the following relation
\begin{equation}
        \alpha_{2.5} = \Delta m_c \times \alpha_{m_c}
\end{equation}
with 
\begin{equation}
        \Delta m_c = \frac{\int_{m^{\star}}^{m^{\star} + 2.5}N(m)\mathrm{d}m}{\int_{m^{\star}}^{m_c}N(m)\mathrm{d}m},   
        \label{eqn:extra-alpha}
\end{equation}
derived from Eq. \eqref{eqn:alphalim}.

In Sect. \ref{sec:PNLF} we derive the PNLF of M49, which we use here as $N(m)$. For $m_c = m^{\star} + 2 = m_2$, the extrapolation factor to $\alpha_{2.5}$ is $\Delta_2 = 1.6$, which results in the following parameters:
\begin{itemize}
        \item $\alpha_{2.5,\mathrm{inner}}^{\mathrm{M49}} = (3.20\pm 0.43) \times 10^{-9}\,\mathrm{PN}\,L^{-1}_{\odot,\mathrm{bol}}$ for the inner component, and
        \item $\alpha_{2.5,\mathrm{outer}}^{\mathrm{M49}} = (1.03\pm 0.22) \times 10^{-8}\,\mathrm{PN}\,L^{-1}_{\odot,\mathrm{bol}}$ for the outer component.
\end{itemize}
For comparison, \citet{1990ApJ...356..332J} determine $\alpha_{2.5} = (6.5\pm 1.4) \times 10^{-9}\,\mathrm{PN}\,L^{-1}_{\odot,\mathrm{bol}}$ for the very inner regions of M49. 

We find that the outer halo has a $\alpha$-parameter that is 3.2 times higher compared to the inner halo. This is suggestive of a change of the parent stellar population with radius. This might be due to a contribution of intragroup PNe from the subcluster B to the outer halo, as was observed for M87 in Virgo's subcluster A \citep[\citetalias{2013A&A...558A..42L};][]{2015A&A...579A.135L}. 

Combining the two-component PN number density model with the colour profile from \citet{2013ApJ...764L..20M}, we were able to evaluate the relation between $\alpha$-parameter and $(B-V)$ colour continuously for the entire galaxy. Motivated by the two-component photometric model described earlier in this section, we extended this approach to the colour profile. At each radius the colour was calculated as the luminosity-weighted average from each component, whose spatial distribution is given by the PN surface-density decomposition. We allowed for a radially dependent colour profile in the inner halo that has the form
\begin{equation}
        (B - V)_{\mathrm{inner}}(r) = \Delta_{\mathrm{i}}r + (B - V)_{\mathrm{i}},
\end{equation} 
where the best fit results in a colour gradient of $\Delta_{\mathrm{i}} = (-1.0 \pm 0.4)\times 10^{-4}\;\mathrm{mag}\;\mathrm{arcsec}^{-1}$ and $(B - V)_{\mathrm{i}} = (B - V)_{\mathrm{inner}}(r = 0) = 0.96\pm0.01$ with a constant outer colour of $(B - V)_{\mathrm{outer}} = 0.25 \pm 0.11$. The resulting total colour profile is in agreement with the new extended photometry of \citet{2017ApJ...834...16M}. The bluer colour of the IGL is in line with younger, more metal-poor IGL components identified in cosmological simulations \citep{2014MNRAS.437..816C}.

\begin{figure}
\begin{center}
        \includegraphics[width = 8.8cm]{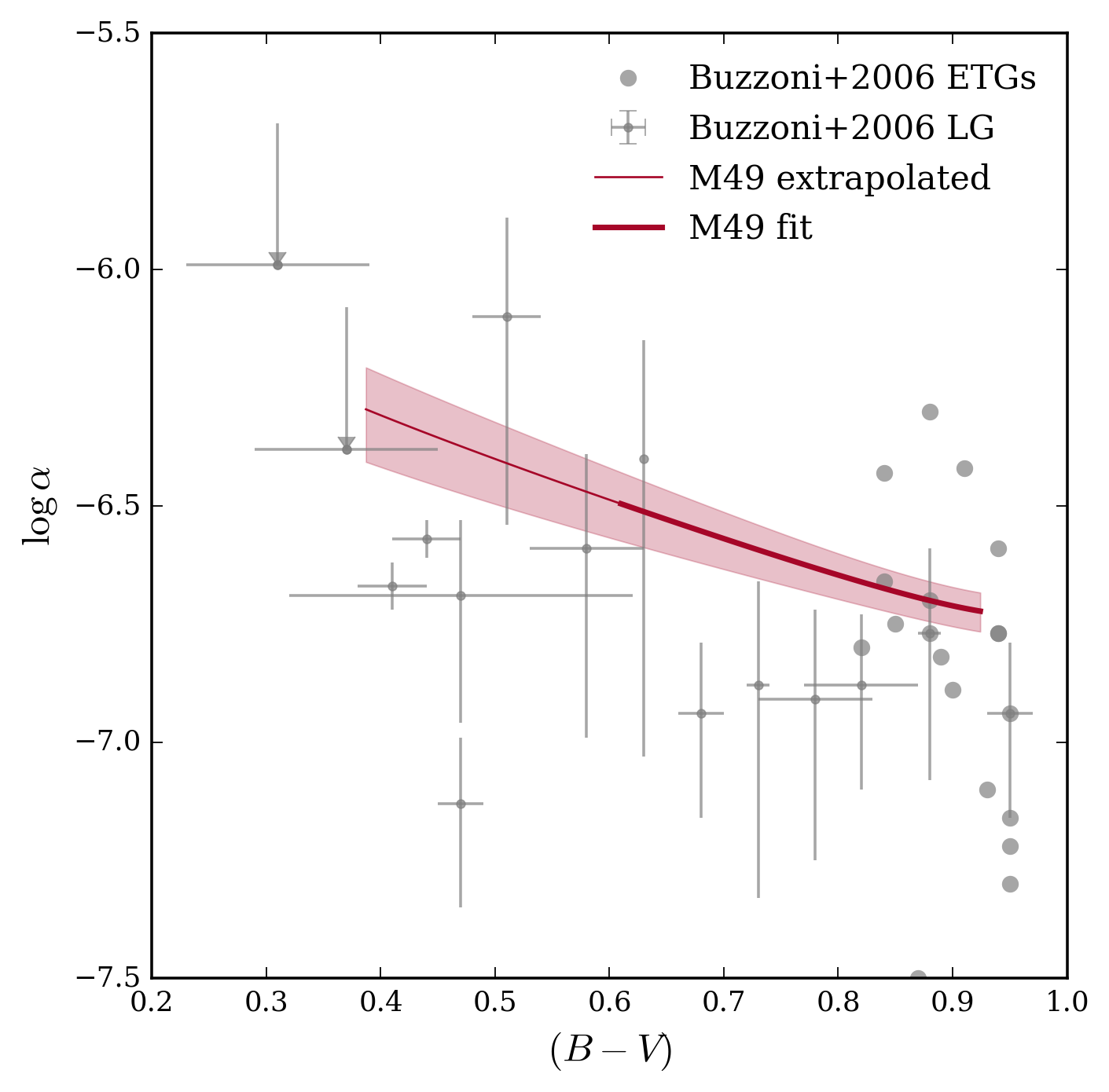}
        \caption{Variation of the $\alpha$-parameter in M49 with colour \citep[red line, colour data from][]{2013ApJ...764L..20M} compared to data from previous PN surveys in Local Group galaxies (grey error bars) and ETGs (grey circles) based on \citet{2006MNRAS.368..877B} and references therein. The result for M49 is based on the two-component model presented in Sect. \ref{ssec:2comp}. The thinner line denotes data exceeding the radial range of the colours measured by \citet{2013ApJ...764L..20M} where the colours have been extrapolated based on the model in Sect. \ref{ssec:2comp}.}
        \label{fig:alpha-colour}
\end{center}
\end{figure}

Figure \ref{fig:alpha-colour} shows the relation for the modelled $\alpha$-parameter as a function of galaxy colour computed from the above model compared to data points of individual galaxies from previous PN surveys of early-type and Local Group (LG) galaxies collected by \citet[][and references therein]{2006MNRAS.368..877B}. We scale our $\alpha$-parameters from $\alpha_{2.5}$ to $\alpha_{8}$ using Eq. \ref{eqn:extra-alpha} based on the best-fit PNLF (Sect. \ref{sec:PNLF}) in order to use the same reference system as \citet{2006MNRAS.368..877B}. As our survey spans a radial range that is $\sim 500^{\prime\prime}$ larger compared to \citet{2013ApJ...764L..20M}, the colour profile has been extrapolated from $1260^{\prime\prime}$ to $1785^{\prime\prime}$ based on our best-fit model. The relation between $\alpha$ and $(B-V)$ colour can be approximated with the following second order polynomial:
\begin{equation}
\alpha(B-V) = 0.34(B-V)^2 -1.25(B-V) -5.86
\label{eqn:alpha-colour}
\end{equation} 
The $\alpha$-parameter is lowest in the inner, red, metal-rich halo, as it would be empirically expected from bright ETGs. It increases with decreasing colour and is highest in the metal-poor, blue, outer halo, similar to the star-forming LG galaxies with this colour.

\subsection{Comparison to GCs in M49's halo}  

\begin{figure}
\begin{center}
        \includegraphics[width = 8.8cm]{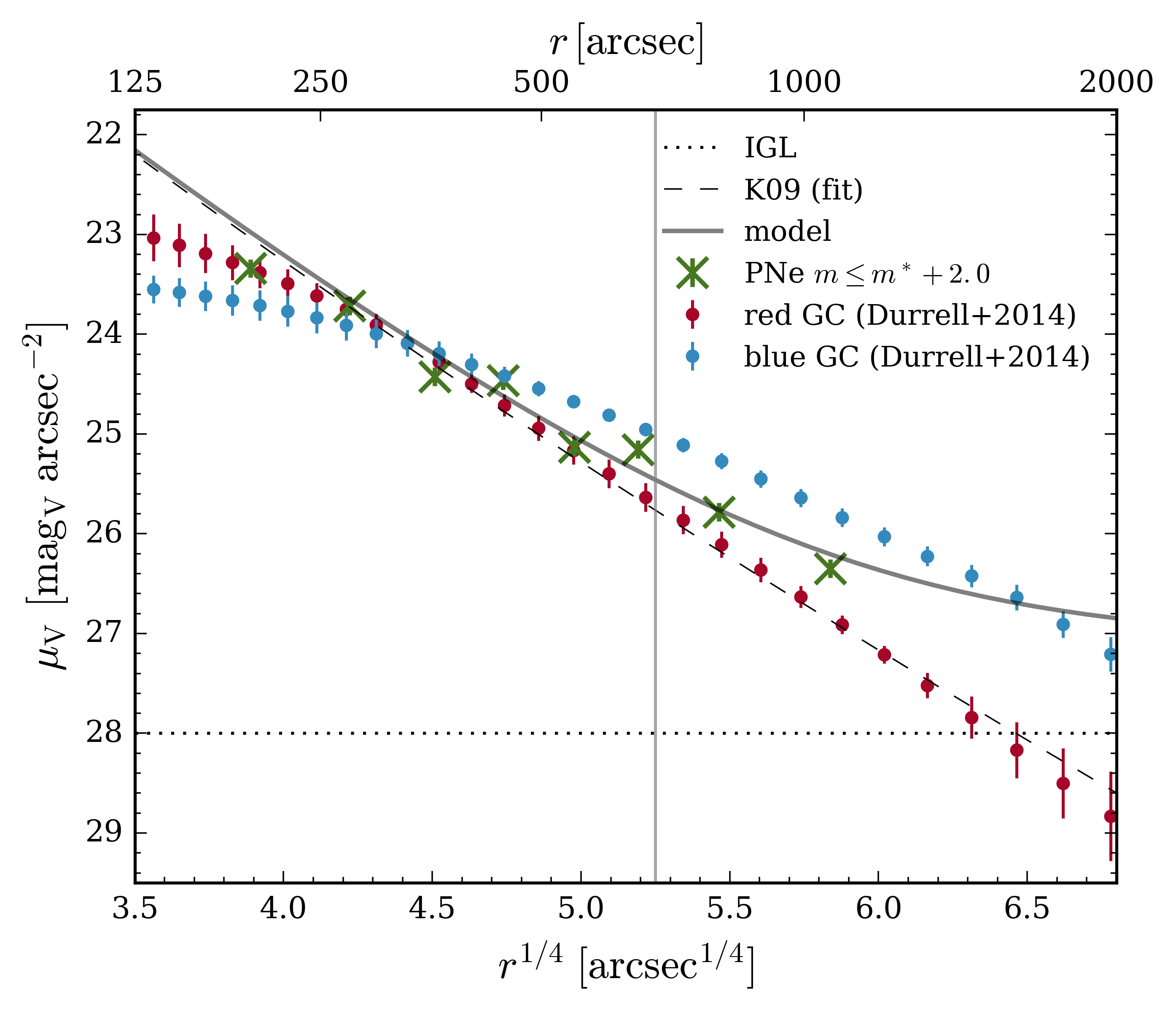}
        \caption{Comparison of the PN number density profile with that of red and blue globular clusters. The S\'ersic fit to broad-band photometry \citepalias{2009ApJS..182..216K} is indicated by the dashed line. The dotted line shows expected contribution of the IGL. The PN number density profile is shown with green crosses and the two-component model fit to it is denoted with the solid grey line. The density profile from red and blue globular clusters are indicated by filled circles in the respective colours. The vertical line denotes the transition radius from the inner to the outer halo based on the change in the PN number density slope.}
        \label{fig:GCs}
\end{center}
\end{figure}

GCs are another tracer of faint extended galaxy halos. The Next Generation Virgo Survey \citep[NGVS,][]{2012ApJS..200....4F} covers the subclusters A and B out to their virial radii, being sufficiently deep to detect GCs \citep{2014ApJ...794..103D}. The GC density profiles shown in Fig. \ref{fig:GCs} are obtained through isophote-fitting of the smoothed red (RGC) and blue (BGC) distributions. In order to compare with the PN number density, we match the distribution of the RGCs to the stellar light in the inner halo \citepalias{2009ApJS..182..216K}. The number density of the PNe and the RGCs agree out to a radius $r^{1/4} = 5.5$, which is where the PNe also start to deviate from the stellar light. The density profile of the BGCs is flatter than both the stellar light and the PN density over the entire radial range covered by both surveys.

\section{The planetary nebula luminosity function of M49}
\label{sec:PNLF}
\begin{figure*}
\begin{center}
        \includegraphics[width = 8.8cm]{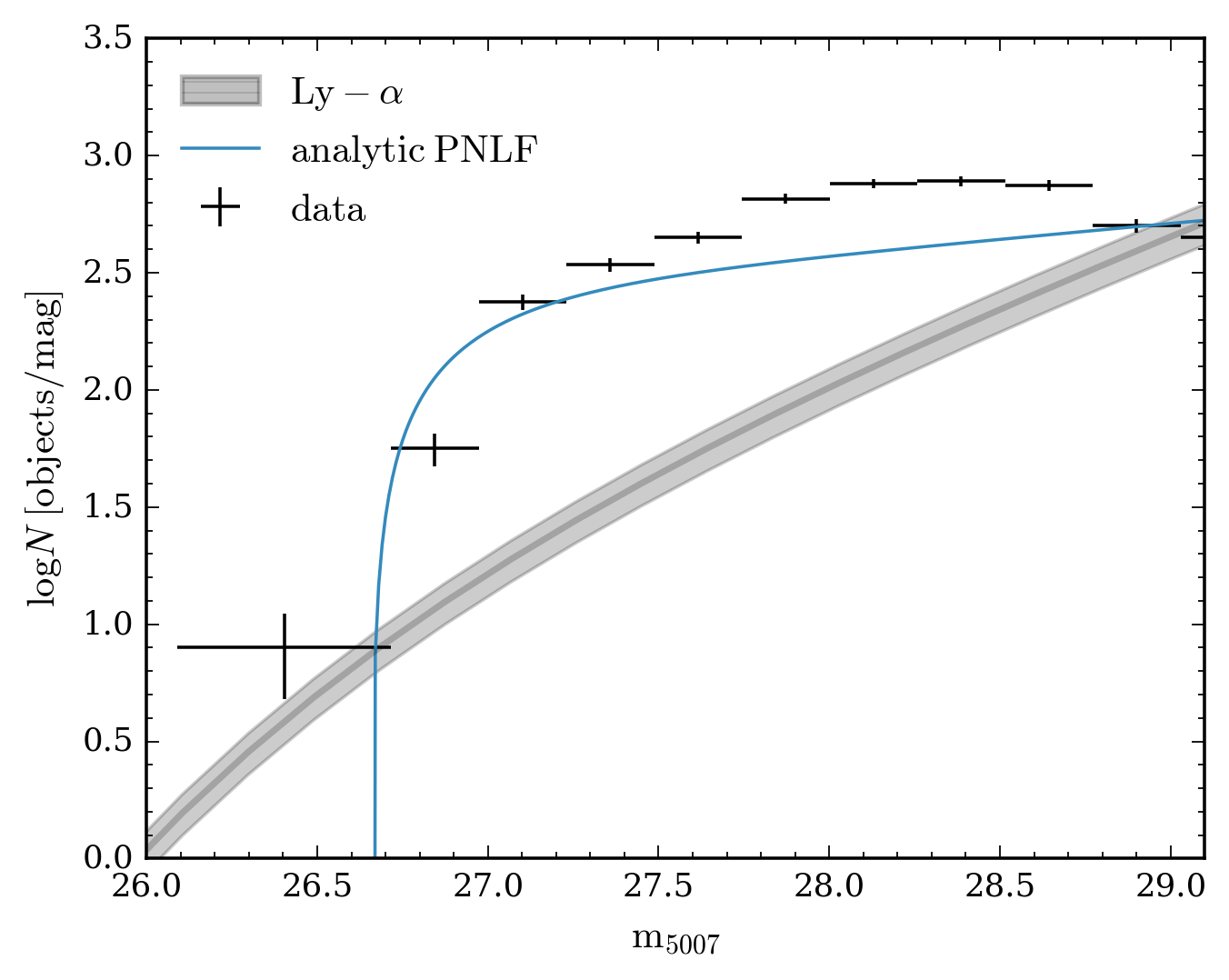}
        \includegraphics[width = 8.8cm]{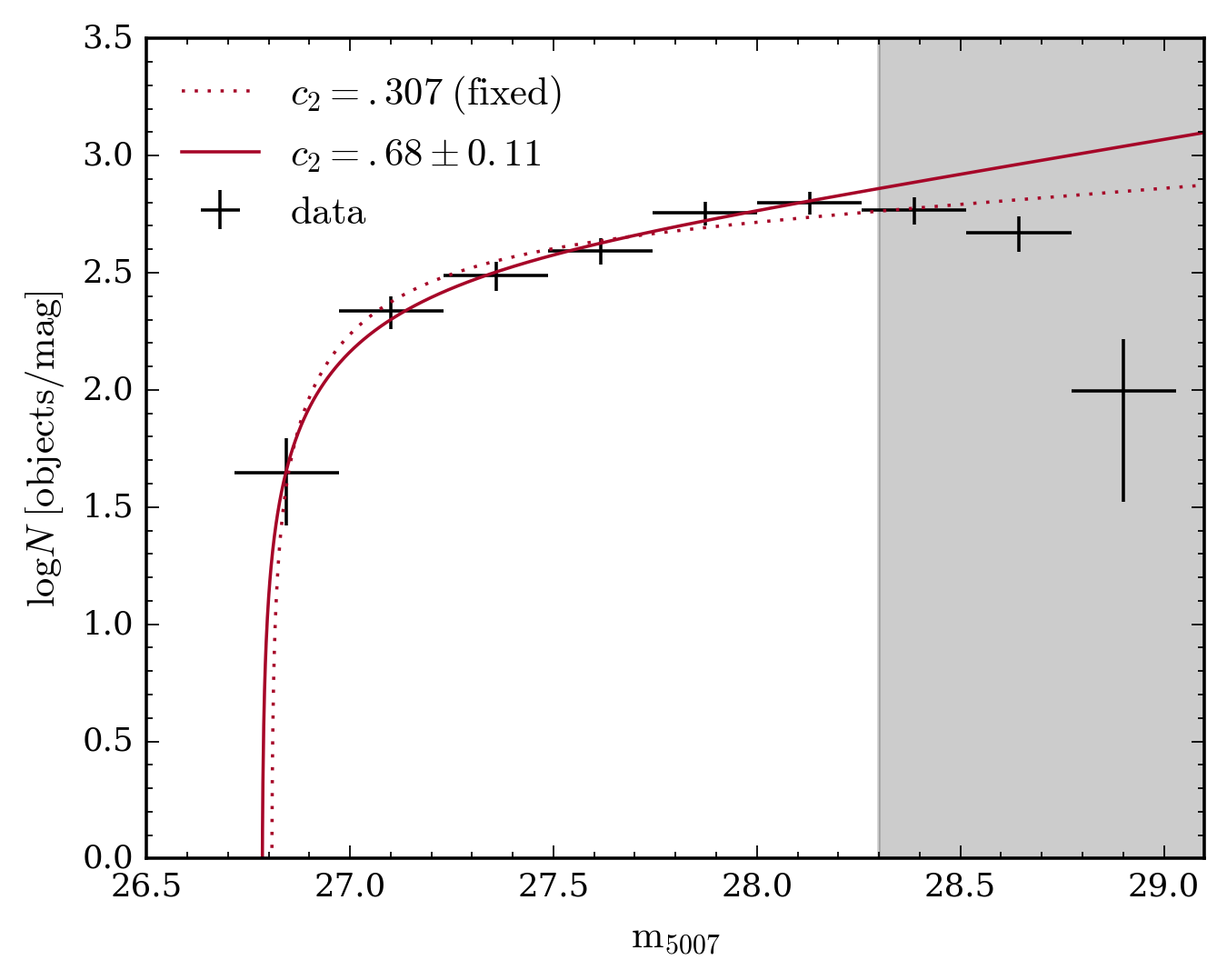}
        \caption{\textit{Left:} PNLF of the completeness-corrected PN sample (without accounting for contamination by Ly-$\alpha$-emitting background galaxies). The error bars along the y-axis account for the Poissonian error and the x-errors for the bin width. The solid blue line shows the analytical PNLF at a distance modulus $\mu = 31.3$ with $c_2 = 0.307$ \citep{1989ApJ...339...53C}. The solid grey line shows the Ly-$\alpha$ LF \citepalias{2007ApJ...667...79G} and its variance due to density fluctuations is represented by the shaded region. \textit{Right:} PNLF of the completeness-corrected PN sample after statistical subtraction of the estimated Ly-$\alpha$ LF. The grey region denotes the magnitude range in which the PNLF might be affected by incompleteness that could not be corrected for. The error bars along the y-axis account for the Poissonian error and the x-errors for the bin width. The solid red line shows the fit of the generalised PNLF, with the best-fit parameters $c_2 = 0.69\pm0.12$ and $\mu = 31.29\pm0.05$. The dotted line shows the fit to the PNLF for $c_2 = 0.307$ \citep{1989ApJ...339...53C}.} 
        \label{fig:PNLF_nocorr}
\end{center}
\end{figure*}

Another characteristic of a PN sample is the shape of its luminosity function (PNLF). As there is empirical evidence for a bright cut-off of the luminosity function, which is consistent with a nearly constant absolute magnitude, the PNLF can be used as a secondary distance indicator \citep{2002ApJ...577...31C}. This cut-off is found to be invariant between different Hubble types \citep{2004ApJ...614..167C}.  
We compared the PNLF of our completeness-corrected PN sample to the generalised analytical formula introduced by \citetalias{2013A&A...558A..42L}
\begin{equation}
        N(M) = c_1 e^{c_2 M} (1 - e^{3(M^{\star} - M)}),
        \label{eqn:PNLF}
\end{equation}
where $c_1$ is a normalisation constant, $c_2$ is the slope at the faint end, and $M^{\star}$ is the absolute magnitude of the LF's bright cut-off. The \citet{1989ApJ...339...53C} analytical LF is then a specific case of Eq. \eqref{eqn:PNLF} with $c_2 = 0.307$ that reproduces the best fit to the PNLF of M31. It describes the exponential envelope expansion \citep{1963ApJ...137..747H} combined with a sharp exponential truncation at the bright end. Observations also suggest that the slope described by the parameter $c_2$ is correlated with the star formation history of the parent stellar population. Steeper slopes are associated with older stellar populations and conversely flatter slopes with younger populations \citep[\citetalias{2013A&A...558A..42L};][]{2004ApJ...614..167C, 2010PASA...27..149C}.  

The left panel of Fig. \ref{fig:PNLF_nocorr} shows the PNLF of the data corrected for detection and colour incompleteness before having subtracted the contribution from Ly-$\alpha$ emitting galaxies together with the expected LF of these objects \citepalias[][]{2007ApJ...667...79G}. The panel also shows the PNLF expected for a galaxy at the distance of M49 \citep{2007ApJ...655..144M} based on Eq. \eqref{eqn:PNLF}, with $c_2 = 0.307$, normalised to the number of objects in our sample. Furthermore, compared to the bright cut-off of the expected PNLF, there are a number of overluminous objects, whose nature we will discuss in the next section.

\subsection{The PNLF distance to M49}
We calculated the PNLF of our PN catalogue, corrected it for detection and colour incompleteness as a function of magnitude (see Sect. \ref{ssec:phot_compl} for details), and statistically subtracted the expected number of Ly-$\alpha$ emitters in each magnitude bin based on the \citetalias{2007ApJ...667...79G} LF. The right panel of Fig. \ref{fig:PNLF_nocorr} shows the resulting PNLF with black crosses indicating the bin width, and the errors due to counting statistics and the uncertainty of the Ly-$\alpha$ LF. Due to the sharp decrease of the PNLF beyond $m_{5007} = 28.8$, we adjust our limiting accordingly, as discussed in Sect. \ref{ssec:vis_insp}. As the generalised function does not account for a decrease of the PNLF with magnitude, we fitted the PNLF to a maximal magnitude of $m_{\mathrm{max}} = 28.3$.

For a fixed parameter $c_2$, we determined a distance modulus $\mu = 31.34 \pm 0.04$ with $\chi^2_{\mathrm{red}} = 1.3$. Allowing for a variation of the slope we fit $c_2 = 0.69\pm0.12$ and $\mu = 31.29\pm0.05$ with $\chi^2_{\mathrm{red}} = 0.4$. A PNLF with a steeper slope compared to M31 \citep{1989ApJ...339...53C} is clearly favoured, but the distance modulus determined in the two fits is similar. Besides of the error in the best-fit, we also needed to account for further sources of error. The random uncertainty in the determination of the photometric zero-point is $\Delta\mathrm{Z_{[OIII]}} = 0.04$. The absolute bright cut-off of the PNLF is $M^{\star} = -4.51^{+0.02}_{-0.04}$ \citep{2002ApJ...577...31C}. Our final distance to M49 is thus $\mu_{\mathrm{PNLF}} = 31.29^{+0.07}_{-0.08}$, which is in agreement with the distance modulus determined from surface brightness fluctuations (SBFs), which is $\mu_{\mathrm{SBF}} = 31.17\pm0.07$ \citep[SBFs,][]{2007ApJ...655..144M}. However, it does not agree with an earlier study of PNe in the centre of M49, which determines the distance modulus to be $\mu_{\mathrm{PNLF}} = 30.70\pm0.14$ \citep{1990ApJ...356..332J}. As there is only a very small overlap between this survey and ours, we are not able to conclude whether this difference in distance modulus is to be ascribed to a zero-point offset between the two photometric systems. 

This leads us to the discrepancy between SBF and PNLF distances. Previous studies have generally found a discrepancy between SBF and PNLF distances, which amounts to a mean difference of $\Delta \mu (\mathrm{PNLF} - \mathrm{SBF}) \approx -0.3$ \citep{2002ApJ...577...31C, 2012Ap&SS.341..151C, 2016arXiv161008625M}. SBF, as the PNLF, is a secondary distance indicator that makes use of luminosity fluctuations due to counting statistics in the individual pixels of CCDs whose amplitude is inversely proportional to the galaxy's distance \citep{1988AJ.....96..807T}. Both SBF and PNLF distances are calibrated based on Cepheid distances \citep{2002ApJ...577...31C, 2001ApJ...546..681T}. Several explanations for the discrepancy are discussed in \citet{2012Ap&SS.341..151C}, arguing that it is likely a combination of multiple small effects like zero-point offsets of the two methods and different extinction due to dust in different galaxy types. 

The offsets in distance moduli neither correlate with absolute galaxy magnitude, nor with galaxy colour, however a small correlation is found between offset and SBF distance modulus \citep{2002ApJ...577...31C}. There are multiple factors that could be responsible for this correlation. It could be due to uncertainties in the measurement in relatively nearby spirals \citep[$\mu_{\mathrm{SBF}} \sim 29 -30$,][]{2012Ap&SS.341..151C}, or to contamination by foreground intracluster stars or background galaxies \citep{2002ApJ...577...31C}. As we detail in Sect. \ref{ssec:lyalpha}, Ly-$\alpha$ emitters at redshift $z = 3.1$ can mimic the unresolved [OIII] emission of PNe and cannot be distinguished from PNe solely based on photometric surveys. This contamination affects galaxies with distances moduli larger than $\mu_{5007} \geq 30.5$, and can bias the empirical PNLF towards brighter magnitudes due to the presence of Ly-$\alpha$ emitters brighter than the PNLF cut-off. By comparing the empirical PNLF with the contribution of the Ly-$\alpha$ LF scaled by filter area and depth, the overluminous sources in the left panel of Fig. \ref{fig:PNLF_nocorr} are clearly consistent with Ly-$\alpha$-emitting background galaxies. Further validation must await the spectroscopic follow-up of these objects.

A recent MUSE study of the star-forming disc galaxy NGC 628 drew the attention to another source of contamination: supernova remnants, that also appear as unresolved [OIII]-emitting objects \citep{2016arXiv161109369K}. However, these objects are predominantly expected in star-forming galaxies as the [OIII] emission from these SN remnants comes from the interaction of the SN blast wave with the circumstellar medium of the progenitor. This is relevant for core-collapse SNe in spiral galaxies which do not occur in ETG-like galaxies like M49 \citep[e.g.][]{1999A&A...351..459C}.

We are aware that the PNLF distance determination in M87 \citepalias{2013A&A...558A..42L} still finds a $0.3$ magnitude discrepancy between the PNLF and SBF distance moduli of M87, also for spectroscopically confirmed PN sample \citep{2015A&A...579A.135L}. As Ly-$\alpha$ emitters were identified in the spectroscopic follow-up, any overluminous Ly-$\alpha$ emitter biasing the PNLF cut-off to brighter magnitudes is expected to have been removed from the PN sample.

\section{Tracing halo variation with PNe}
\label{sec:halo_variation}

In Sect. \ref{sec:density} we found evidence for two different PN populations in the inner and outer halo based on a variation of the $\alpha$-parameter. We want to investigate whether these differences are also reflected in the PNLF. The PNLF does not only serve as a distance indicator, but furthermore provides insights into the underlying stellar populations, which influence its morphology at fainter magnitudes than the bright cut-off. 

\subsection{The PNLF morphology in the inner and outer halo}
\begin{figure*}
\begin{center}
        \includegraphics[width = 8.8cm]{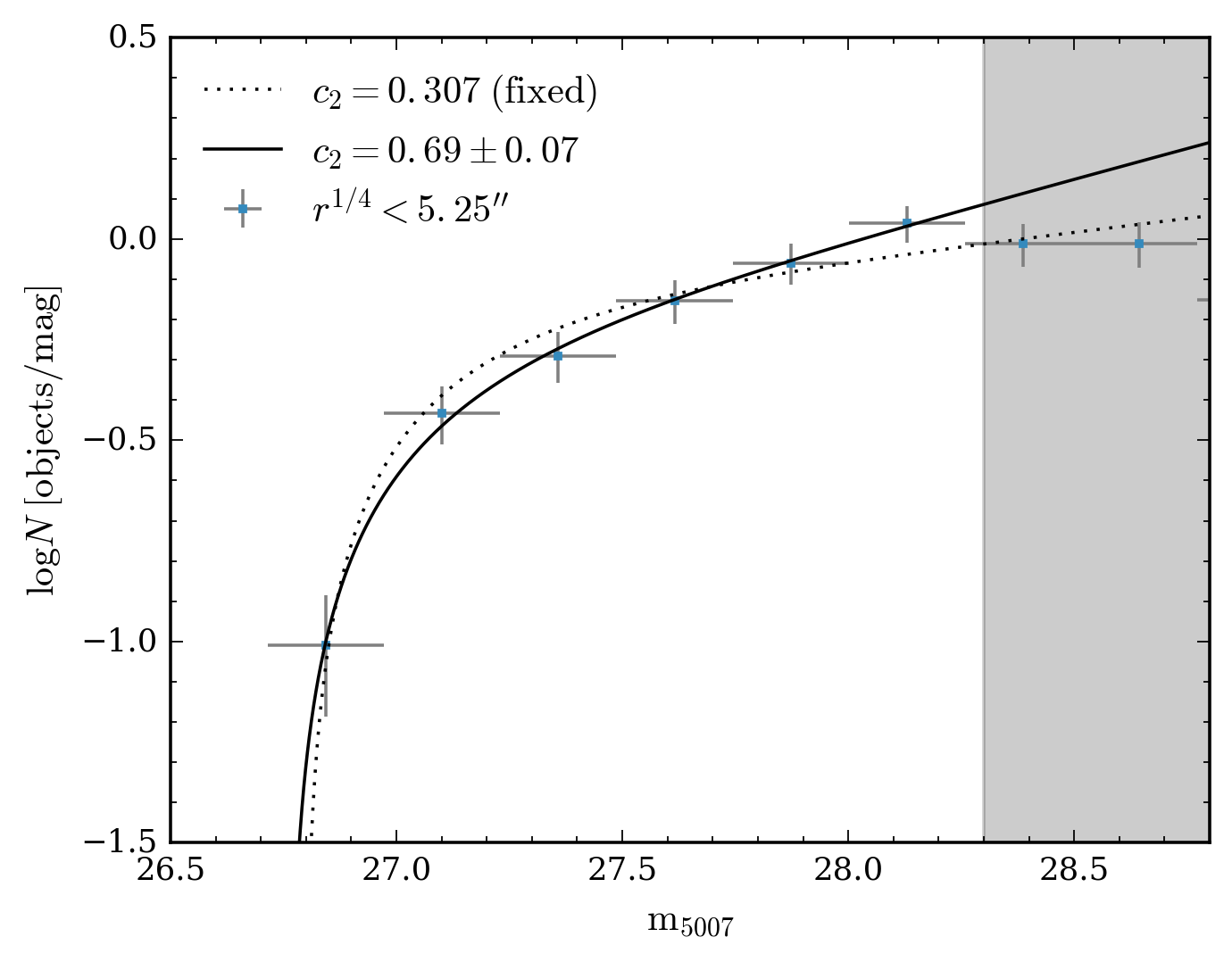}
        \includegraphics[width = 8.8cm]{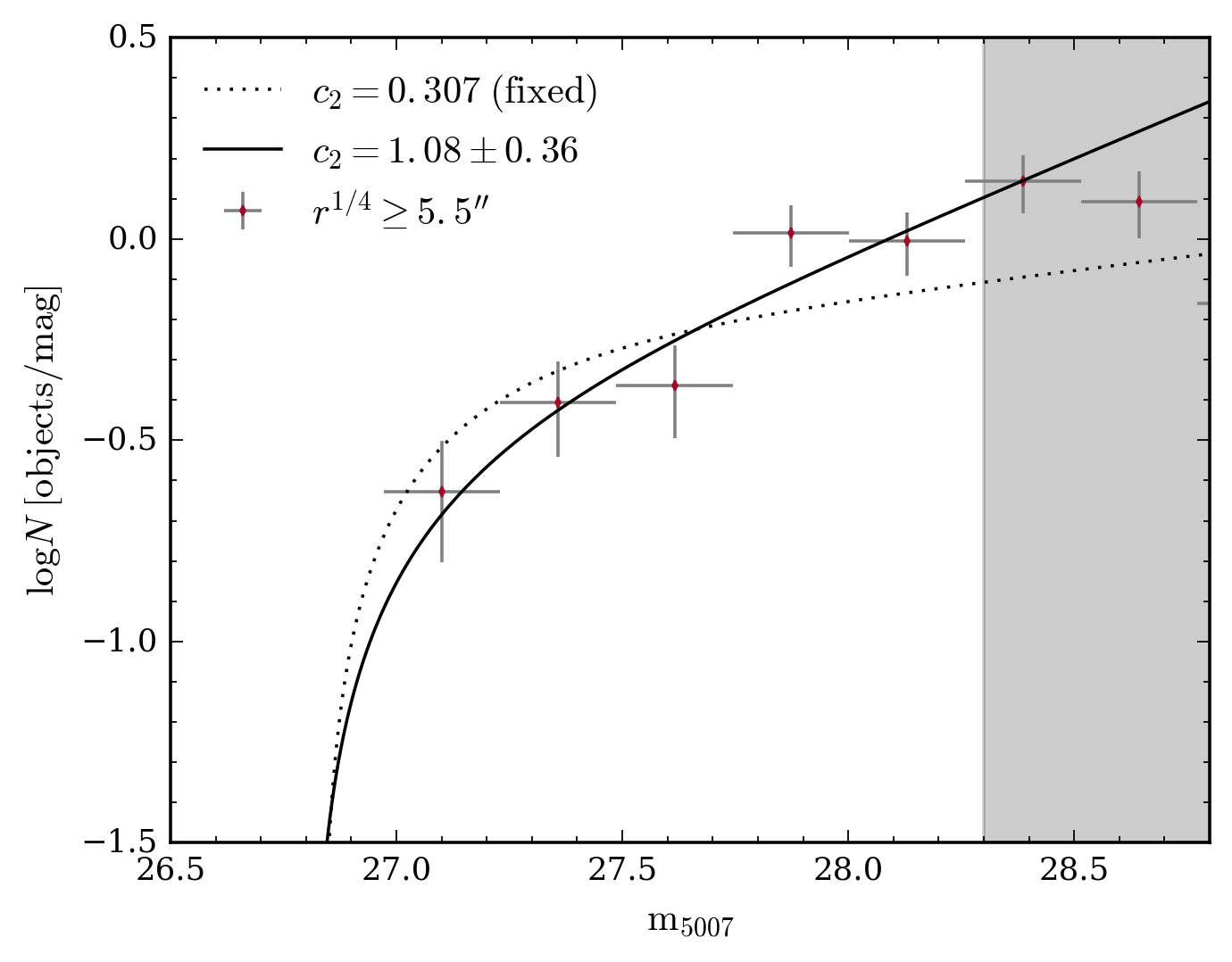}
        \caption{PNLF of the completeness-corrected PN sample in two radial bins. The left panel shows the PN sample in the inner halo ($r^{1/4} < 5.25$) and the right one the sample in the outer halo ($r^{1/4} \geq 5.25$). The error bars along the y-axis account for the Poissonian error and the x-errors for the bin width. The solid lines show the fits to the PNLF allowing for variation in the parameter $c_2$ while the dashed lines show the fits with a fixed $c_2 = 0.307$ \citep{1989ApJ...339...53C}. As in Fig. \ref{fig:PNLF_nocorr}, the grey region denotes the magnitude range in which the PNLF might be affected by incompleteness.}
        \label{fig:PNLFsplit}
\end{center}
\end{figure*}

We first divided the PN sample into two concentric elliptical bins, indicated by the grey line in Fig. \ref{fig:density} at $r^{1/4} = 5.25$. The inner sample is characterised by a density profile that decreases steeply with radius, and contains 517 PNe, whereas the outer sample has a shallower number density profile and contains 197 objects (see also Sect. \ref{sec:density}). We computed the completeness-corrected PNLF in each bin, normalising by the total number of objects for better comparison and again fit the generalised PNLF. The resulting profiles are shown in Fig. \ref{fig:PNLFsplit}. For comparison we also show the PNLF with a fixed $c_2 = 0.307$, which is clearly disfavoured in both bins. The PNLF in the outer bin is less well fit by the generalised function than the one in the inner bin. The fitted distance moduli are $\mu_{\mathrm{PNLF, inner}} = 31.27 \pm 0.02$ and $\mu_{\mathrm{PNLF, outer}} = 31.31 \pm 0.05$, which agree with the distance modulus determined in the previous section.

In the outer regions of the halo, the slope of the PNLF is steeper ($c_{2, \mathrm{outer}} = 1.08 \pm 0.36$) than in the inner region ($c_{2, \mathrm{inner}} = 0.69 \pm 0.07$). Empirically, one would thus expect an older underlying stellar population in the outer halo compared to the inner halo. This seems counterintuitive, given the colour gradient towards bluer colour and the general understanding of halo growth through accretion of smaller, younger systems \citep{2012ApJ...744...63O, 2013MNRAS.434.3348C}.  In the following section we investigate the properties of the PNLF in the inner halo.

\subsection{PNLF variation in the inner halo}
\label{ssec:PNLFvar}
\begin{figure*}
\begin{center}
        \includegraphics[width = 17cm]{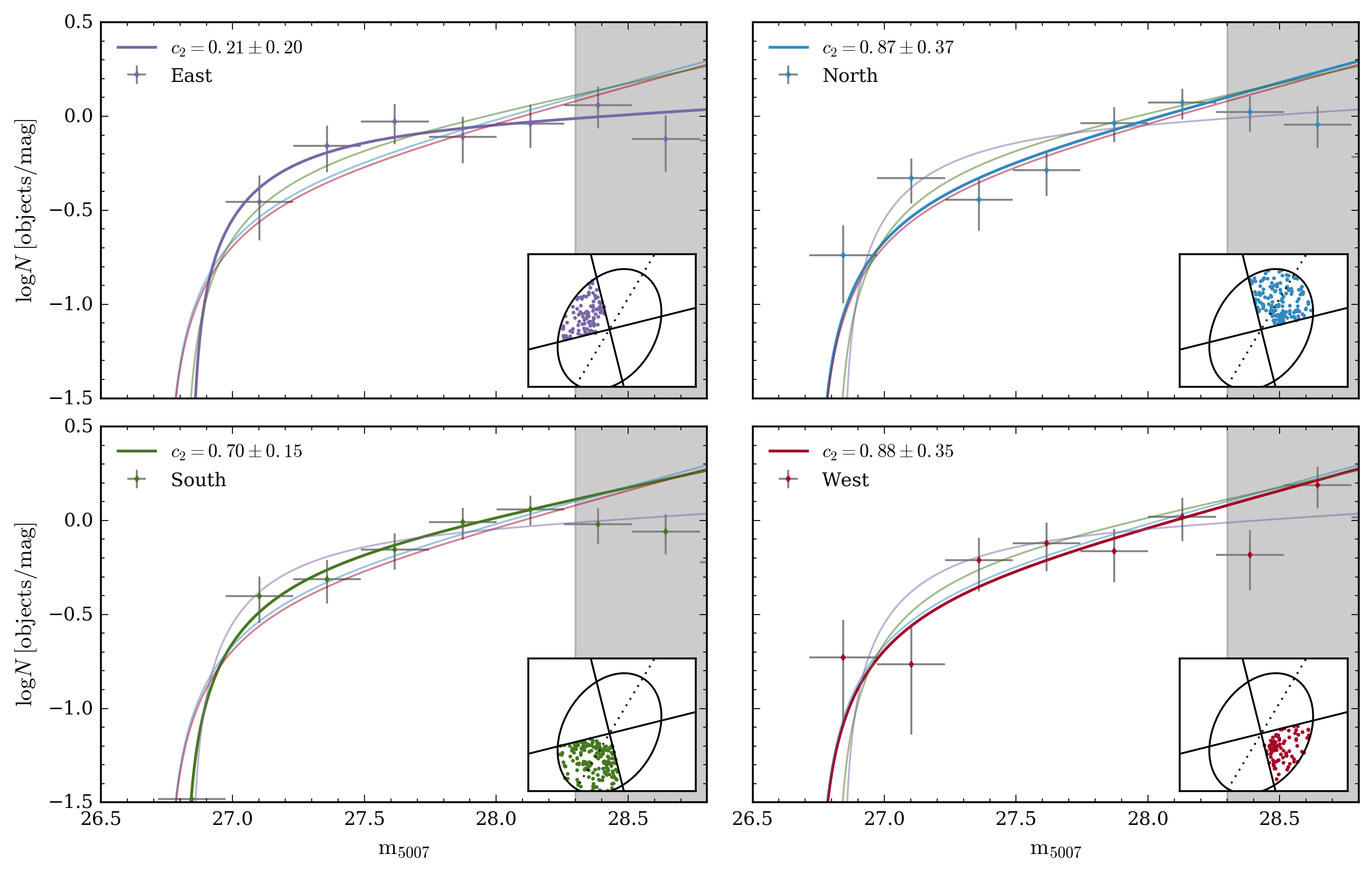}
        \caption{PNLF variation in the inner bin in four quadrants. Each panel shows the observed PNLF (data points and error bars) and the corresponding fit for the quadrant indicated in the inset.}
        \label{fig:cones}
\end{center}
\end{figure*}
As M49's halo is rich in substructure \citep[][\citetalias{2015A&A...581A..10C}]{2010ApJ...715..972J}, we next investigate the azimuthal variation of the PNLF in the inner halo. We therefore divide the halo into four quadrants, such that the positive and negative minor and major axes are enclosed respectively, as shown in Fig. \ref{fig:2Ddensity}. The quadrants are denominated by their position with respect to M49's centre (north, west, east, south). We calculate the completeness-corrected PNLF in each quadrant and normalise it by the total number of PN in that quadrant. The resulting profiles are shown in Fig. \ref{fig:cones}. We again fitted the generalised PNLF to each profile.

While the PNLF in the northern, southern, and western quadrants have similar slopes ranging from $c_2 = 0.87 \pm 0.37$ in the north to $c_2 = 0.70 \pm 0.15$ in the south, the slope of the PNLF in the eastern quadrant is significantly flatter, being $c_2 = 0.21 \pm 0.20$, even shallower than the slope of the Ciardullo-PNLF. Compared to the other quadrants, the bright end of the PNLF in the east is populated by more objects. This drives the flattening of the PNLF in the inner halo, which we described in the previous section. If this flattening is related to a younger stellar population, it likely points towards an accretion event. No substructure is found in deep broad-band photometric studies \citep{2010ApJ...715..972J} in this quadrant, which could be associated to the PNLF variation. However, the spatial proximity to the trail of the accreted dIrr galaxy VCC 1249, which could well extend further into the eastern quadrant, argues for a relation between the flat PNLF and this accretion event.

\subsection{2D number density map}
\label{ssec:2Ddensity}

\begin{figure}
\begin{center}
        \includegraphics[width = 8.8cm]{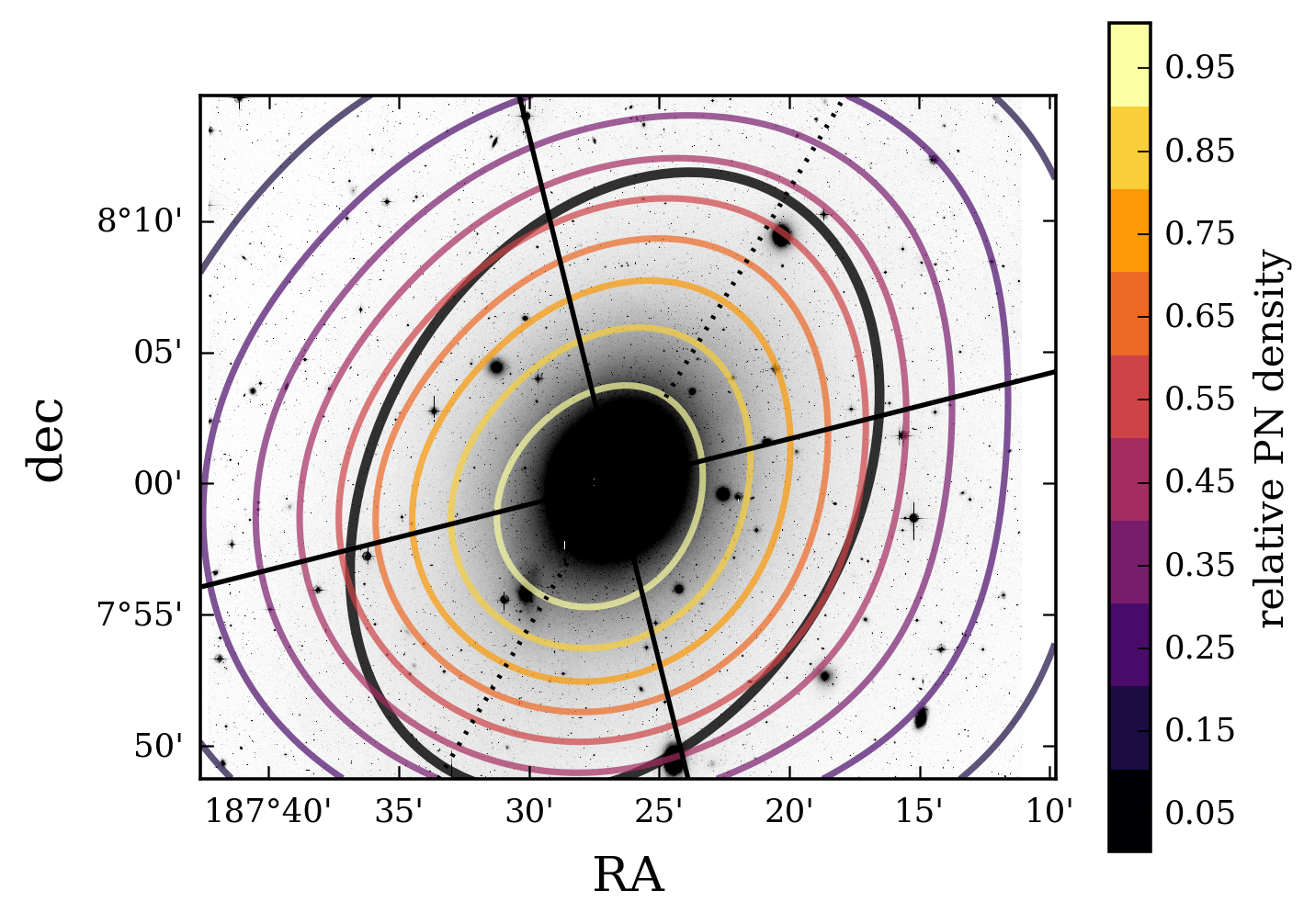}
        \caption{NGVS $g$-band image of M49 \citep[from][]{2012A&A...543A.112A} with 2D contours of the relative PN density superimposed. The black ellipse denotes the major-axis transition radius from the inner to the outer halo at $730''$ as defined in Sect. \ref{sec:density}, the dashed line indicates the major axis and the solid lines the borders of the quadrants as defined in Sect. \ref{sec:halo_variation}.}
        \label{fig:2Ddensity}
\end{center}
\end{figure}

We next evaluated whether we can identify any substructure in the 2D PNe density distribution. We therefore performed a kernel density estimate (KDE) with a Gaussian kernel using \texttt{scikit-learn} \citep{scikit-learn}. The kernel bandwidth was chosen such that no artificial overdensities are introduced due to the masked regions in the centre. The resulting density contours overplotted on the NGVS $g$-band image are shown in Fig. \ref{fig:2Ddensity}. This Fig. also shows the borders of the quadrants defined in the previous Sect., as well as the boundary between the inner and outer halo. Due to the large bandwidth we were not able to resolve individual structures. The PN-contours along the north of the major axis have a similar position angle and ellipticity as the isophotes from broad-band studies, in the south are flattened and denser. This spatial squashing coincides with region of interaction with VCC 1249. 

\section{Discussion}
\label{sec:discussion}
Hydrodynamical cosmological simulations such as the Illustris simulations \citep{2014MNRAS.444.1518V} predict that accretion and mergers will deposit significant amounts of tidally stripped stars in the outer regions of galaxies \citep{2015ApJ...799..184P}. Because the dynamical timescales in galaxy outskirts, at 50 – 150 kpc radii, are of the order of 2 Gyr, this predictions can be tested by looking observationally for unrelaxed dynamical structures at these radii. This motivates the investigation in this paper of the observational evidence for substructures, environmental effects and intragroup light around the giant Virgo galaxy M49.

\subsection{Substructure in M49}
\label{ssec:substructure}
Recent deep broad-band photometric studies have revealed an intricate system of stellar substructure in M49's halo. \citet{2010ApJ...715..972J} fit S\'ersic and double-deVaucouleurs profiles to their $V$-band photometry and subtract the resulting galaxy model from the images in order to identify substructures. They find a system of diffuse shells, mainly along the major axis of the galaxy. Furthermore, there are two plumes to the south and in the north-east of the galaxy, the latter running through the dwarf galaxy VCC 1254. There might be a connection between the south plume and the inner shell system. This plume is also closely located to the dwarf irregular (dIrr) galaxy VCC 1249, whose properties will be discussed later in this section. The colours of the substructures that were identified by \citet{2010ApJ...715..972J} were measured by \citet{2013ApJ...764L..20M}. Their measurement of the global colour trend in the halo of M49 was already discussed in Sect. \ref{sec:density}. The starforming companions of M49 (VCC 1249 and VCC 1205) clearly stand out due to their bluer $B - V$ colour. The inner shells cannot be identified in the colour map, as they are too close to the bright and red main halo of the galaxy. However, the extended debris shell in the north-west can be identified in the colour map. It is $\sim 0.07$ mag redder than the remainder of the halo at this radius, indicating that the shell is a remnant of a disruptive accretion event. 

The interaction between the dIrr galaxy VCC 1249 and M49 has come to the attention after the detection of an H \textrm{I} cloud displaced from the dwarf towards M49 \citep{1983AJ.....88..260K}. Follow-up VLA observations of this cloud by \citet{1994AJ....108..844M} revealed a debris trail between the dIrr and M49. A multi-wavelength analysis of this region \citep{2012A&A...543A.112A} including near-UV data as well as optical imaging, H$\alpha$ and spectroscopy comes to the conclusion that star formation in VCC 1249 has been quenched. Due to ram pressure stripping and tidal interaction H \textrm{I} gas has been removed from the dIrr core. Star formation remains active in the H \textrm{I} cloud. 

Spatially coinciding with this accretion event, we observe denser PN contours south-east of M49's centre (Sect. \ref{ssec:2Ddensity}). Having evaluated the PNLF in different quadrants, we however do not find a significant change in the PNLF morphology in this region (Sect. \ref{ssec:PNLFvar}). A significant difference in the PNLF slope is indeed found in the adjacent eastern quadrant. The flattening of the PNLF slope can have multiple origins, being driven by either an overabundance of bright to medium luminosity PNe or a lack of fainter PNe. Based on Fig. \ref{fig:cones} we assume that the presence of additional medium luminosity PNe $(27.2 \leq m_{5007} \leq 27.7)$ compared to the general halo population is driving the flattening of the PNLF. The presence of these bright objects could very well be related to the accretion of VCC 1249, as the shedding of tidal debris after its pericentre passage can lead to a wider distribution of stars (and therefore PNe) than inferred from the distribution of the light \citep[e.g.,][]{2015MNRAS.450..575A}.
The validation of this claim will be one of the aims of the spectroscopic follow-up survey of this data, with which we expect to be able to identify accretion events based on their signatures in the kinematic phase-space.

\subsection{The effect of environment: a comparison to M87}
\label{ssec:M87comp}
M87 has been a milestone for this type of PN-surveys in the Virgo cluster. With the spectroscopic follow-up of the deep photometric PN-survey \citepalias{2013A&A...558A..42L}, \citet{2015A&A...579A.135L} were able to kinematically separate the stellar halo and intracluster light and furthermore found evidence for a late accretion event in M87's velocity phase-space \citep{2015A&A...579L...3L}.  

We first compared the properties of the PN number density profiles in the two galaxies. The morphology of the number density profiles is very similar, both closely following the stellar light \citepalias{2009ApJS..182..216K} in the inner galaxy halos and then flattens at larger radii. In both galaxies, a two-component model is fit that allows for different $\alpha$-parameters of the galaxy PN and the PN associated with the ICL/IGL. We scaled M87's $\alpha$-parameters by a factor of $0.54$, as they were calculated assuming a distance modulus of $\mu_{5007} = 30.8$ \citepalias{2013A&A...558A..42L} instead of $\mu_{\mathrm{SBF}} = 31.17$ \citep{2007ApJ...655..144M}, resulting in 
\begin{itemize}
        \item $\alpha_{2.5,\mathrm{halo}}^{\mathrm{M87}} = (5.94^{+0.17}_{-0.21}) \times 10^{-9}\,\mathrm{PN}\,L^{-1}_{\odot,\mathrm{bol}}$ for the halo component, and
        \item $\alpha_{2.5,\mathrm{ICL}}^{\mathrm{M87}} = (1.78^{+0.60}_{-0.72}) \times 10^{-8}\,\mathrm{PN}\,L^{-1}_{\odot,\mathrm{bol}}$ for the ICL component \citepalias{2013A&A...558A..42L}.
\end{itemize}
The ratio between the two parameters is $\alpha_{2.5,\mathrm{ICL}}/\alpha_{2.5,\mathrm{halo}} = 3$. In the photometric study, the contribution of the ICL was expected to be constant $\mu_\mathrm{V} = 27.7\;\mathrm{mag}\;\mathrm{arcsec}^{-2}$. In M49, we assume that the contribution from the IGL is $\mu_\mathrm{V} = 28.0\;\mathrm{mag}\;\mathrm{arcsec}^{-2}$. For reference M49's $\alpha$-parameters as determined in Sect. \ref{sec:density} of this paper are 
\begin{itemize}
        \item $\alpha_{2.5,\mathrm{inner}}^{\mathrm{M49}} = (3.20\pm 0.43) \times 10^{-9}\,\mathrm{PN}\,L^{-1}_{\odot,\mathrm{bol}}$ for the inner component (halo), and
        \item $\alpha_{2.5,\mathrm{outer}}^{\mathrm{M49}} = (1.03\pm 0.22) \times 10^{-8}\,\mathrm{PN}\,L^{-1}_{\odot,\mathrm{bol}}$ for the outer component (IGL).
\end{itemize}
The $\alpha$-parameter in the main halo of M87 is higher compared to the one in M49, while the ICL and IGL have consistent $\alpha$-values within the errors.

The halo PNLF of M87 is steeper than what is expected based on the analytic PNLF from \citet{1989ApJ...339...53C}, which motivated in the introduction of the generalised analytic PNLF in \citetalias{2013A&A...558A..42L}. The distance modulus and the slope are fit to be $\mu = 30.73$  and $c_2 = 1.17$. The determined distance modulus is $\sim0.4$ mag brighter than the one determined using SBFs \citep{2007ApJ...655..144M}. In \citet{2015A&A...579A.135L}, the slope of the halo and ICL PNLFs were determined based on the kinematically separated populations, leading to $c_{2,\mathrm{halo}} = 0.72$ and $c_{2,\mathrm{ICL}} = 0.66$. The PNLF of the ICL shows a dip 1 - 1.5 mag from the bright cut-off, which is seen in star-forming systems. There is little variation of the PNLF with radius. 
In contrast to this, we find differences in the PNLF slope in different regions of M49's halo. The slope of the PNLF in the entire sample is also steeper ($c_2 = 0.69\pm0.12$) than the analytic PNLF from \citet{1989ApJ...339...53C}, which further corroborates the empirical trend of steepening PNLFs from star-forming to old metal-rich populations \citep[\citetalias{2013A&A...558A..42L};][]{2004ApJ...614..167C, 2010PASA...27..149C}. We believe that the flat PNLF in the inner halo of M49 is driven by the excess of bright to medium-luminosity planetaries in the eastern quadrant. 

We spatially associate this excess with the tail of the accretion event of VCC 1249, but note that there is no overdensity that can be related with it in the broad-band photometric surveys of M49's halo \citep{2010ApJ...715..972J, 2015A&A...581A..10C}. As identified in \citet{2015A&A...579L...3L}, M87's kinematic PN phase-space reveals signatures of a recent accretion event, also termed as M87's crown. However, the PNLF of the crown in M87 does not show signs of variation from the halo PNLF. The nature of the accreted system is very different from what is observed for the VCC 1249 - M49 interaction. Reflecting the environment of Virgo's subcluster A, the infalling system was devoid of gas. The subcluster B at whose centre M49 resides mainly consists of star-forming galaxies. Thus the infall of a gas-rich, star-forming galaxy is more likely to be observed in the subcluster B. 

\subsection{M49 and its environment: the intragroup light in the Virgo Subcluster B}
\label{ssec:IGL}
Combining our study of individual tracers (PNe) with broad-band photometric observations of M49 and its environment allows us to model the PN populations in M49's halo as a two-component system that consists of an inner S\'ersic-like halo and a flat outer component with a constant surface brightness (see Sect. \ref{ssec:2comp}). Our best-fit model is characterised by the $\alpha$-parameters of the respective components, which we re-stated in Sect. \ref{ssec:M87comp} above.

By applying the same concept to the colour profile \citep{2013ApJ...764L..20M}, we are able to derive a relation between the $\alpha$-parameter and the $(B - V)$ colour at each radius: M49's $\alpha$-parameter increases with decreasing colour. The best-fit colour model consists of an inner component  with a radially dependent colour profile with gradient $\Delta_{\mathrm{i}} = (-1.0 \pm 0.4)\times 10^{-4}\;\mathrm{mag}\;\mathrm{arcsec}^{-1}$ and $(B - V)_{\mathrm{i}} = (B - V)_{\mathrm{inner}}(r = 0) = 0.96\pm0.01$ and an outer component of constant colour $(B - V)_{\mathrm{outer}} = 0.25 \pm 0.11$. The comparison with the most recent hydrodynamical, cosmological simulations, shows that the S\'ersic component and its mild colour gradient of $\nabla (B - V)_{2 - 4r_e} = -0.13 \;\mathrm{mag}\;\mathrm{dex}^{-1}$ are consistent with the hierarchical assembly of early-type galaxies \citep{2016ApJ...833..158C}. In fact, the colour profiles measured in the Illustris simulation have an average $(g - r)$ colour gradient of $\nabla (g - r)_{2 - 4r_e} \approx -0.1 \;\mathrm{mag}\;\mathrm{dex}^{-1}$ \citep{2016ApJ...833..158C} with a scatter of $\sigma \approx 0.08\;\mathrm{mag}\;\mathrm{dex}^{-1}$. This is much shallower than the global gradient observed in M49 that ranges from $\nabla (B - V)_{2 - 4r_e} = -0.3 \;\mathrm{mag}\;\mathrm{dex}^{-1}$ in the inner halo to $\nabla (B - V)_{4 - 8r_e} = -0.8 \;\mathrm{mag}\;\mathrm{dex}^{-1}$ in the outer halo, thus supporting the presence of an additional component. 

The shallow, blue, outer component is likely due to a contribution of a smooth IGL that is superposed onto the S\'ersic galaxy halo. This result is consistent with the extended photometry from the Burrell Schmidt deep Virgo survey  \citep{2017ApJ...834...16M} and the VEGAS survey \citepalias{2015A&A...581A..10C}. Because \citet{2017ApJ...834...16M}'s morphological classification of ICL/IGL is based on the luminosity fraction that is either in filamentary or in clumpy structures, in addition to an azimuthally smooth spheroid, the smooth blue IGL around M49 would not fall in this classification and hence their different conclusion on the presence of IGL around M49. We also note that the footprint of the Burrell Schmidt survey does not extend to the south of M49, where the outer boundary of the external envelope is most elongated \citepalias[see Fig. 6 in][]{2015A&A...581A..10C}. 

As the outer halo of M49 is observed to be smooth \citep[\citetalias{2015A&A...581A..10C}, ][]{2017ApJ...834...16M} we would expect the component to have already started to relaxed. As the precession time at these radii is about $5$ Gyr, following the simple stellar population models of \citet{2003MNRAS.344.1000B} applied to M49 in \citet{2013ApJ...764L..20M}, we expect this population to be quite metal-poor ($\mathrm{[Fe/H]2 < 2}$). Similar to the analysis in M87 by \citet{2015A&A...579A.135L}, we are planning to carry out a spectroscopic follow-up of the PN-survey presented in this paper. This will enable us to assess the degree of relaxation of the outer component, which points towards its evolutionary history. Based on their velocity phase-space properties we will also be able to kinematically disentangle the halo and IGL populations.    

\section{Summary and conclusions}
\label{sec:summary}
We present a deep narrow-band survey of the bright ETG M49 using Subaru's Suprime-Cam in order to identify and analyse the PN population therein. The survey's extent and depth is unprecedented for M49. Down to a limiting magnitude of $m_{5007,\mathrm{lim}} = 28.8$ and covering a radial range of 155 kpc from the galaxy's centre, we identify a PNLF-complete sample of $624$ PNe using automated detection techniques. The selection criteria are based on the fact the PNe have a bright [OIII] and no continuum emission and that they appear as point-like at extragalactic distances. Using a simulated PN population, we are able to account for detection and colour incompleteness. 

The radial PN number density profile follows the broad-band surface brightness distribution to a radius of $r = 730''$, equivalent to $r = 60\;\mathrm{kpc}$ and then shows flattening with respect to the stellar surface brightness profile. This is consistent with a two-component model that accounts for a S\'ersic distribution of  halo PNe in the inner region and a shallower population of IGL PNe at larger radii. The $\alpha$-parameter of this IGL population signifies a 3.2 times higher specific frequency of PNe in the IGL compared to the main galaxy halo, which is similar to what was observed in Virgo's BCG M87 \citepalias{2013A&A...558A..42L}. 

The PNLF is as an important diagnostic tool for the properties of the global PN population of M49's halo as well as for variation therein. First, due to accurate account for contamination by Ly-$\alpha$ emitting background galaxies at $z = 3.1$, we are able to determine a PNLF distance modulus that agrees with the one determined using the SBF technique, which is a first for a galaxy at this distance. Second, the PNLF slope varies in the inner halo of M49, due to additional bright PNe in one quadrant. Whether these PNe are related to the accretion of VCC 1249 will be one of the aims of the planned spectroscopic follow-up. We will also be able to disentangle the halo and intragroup populations by studying their kinematics separately.  

While both M49 and M87 are massive and bright early-type galaxies in the Virgo cluster, their PN populations have different properties, reflecting the different environments in which they reside. M87 in Virgo's most dense environment, the subcluster A, is surrounded by ICL with a higher specific frequency compared to the IGL around M49 in Virgo's subcluster B. Both galaxies are undergoing or have undergone recent accretion. The fact that M87's  accreted companion did not contain any gas, while M49 has multiple star-forming companions reflects the different environments in which the halos of these ETGs grew. 
        
\section*{Acknowledgements}
The authors thank the on-site Subaru staff for their support, Patrick Durrell for the provision of the NGVS number density profiles in M49, and Fabrizio Arrigoni Battaia for the NGVS $g$-band image of M49. 
This research made use of Astropy, a community-developed core Python package for Astronomy \citep{2013A&A...558A..33A}.
This publication made use of data products from the Two Micron All Sky Survey (University of Massachusetts and the Infrared Processing and Analysis Center/California Institute of Technology), funded by the National Aeronautics and Space Administration and the National Science Foundation. 
The authors thank the referee for the constructive comments and the careful reading of the manuscript.

\bibliographystyle{aa} 
\bibliography{M49}

\appendix
\section{Catalogue extraction}
\label{app:selection}

\subsection{Extraction of point-like emission-line objects}
\label{app:extraction}
\begin{figure}
\begin{center}
        \includegraphics[width = 8.8cm]{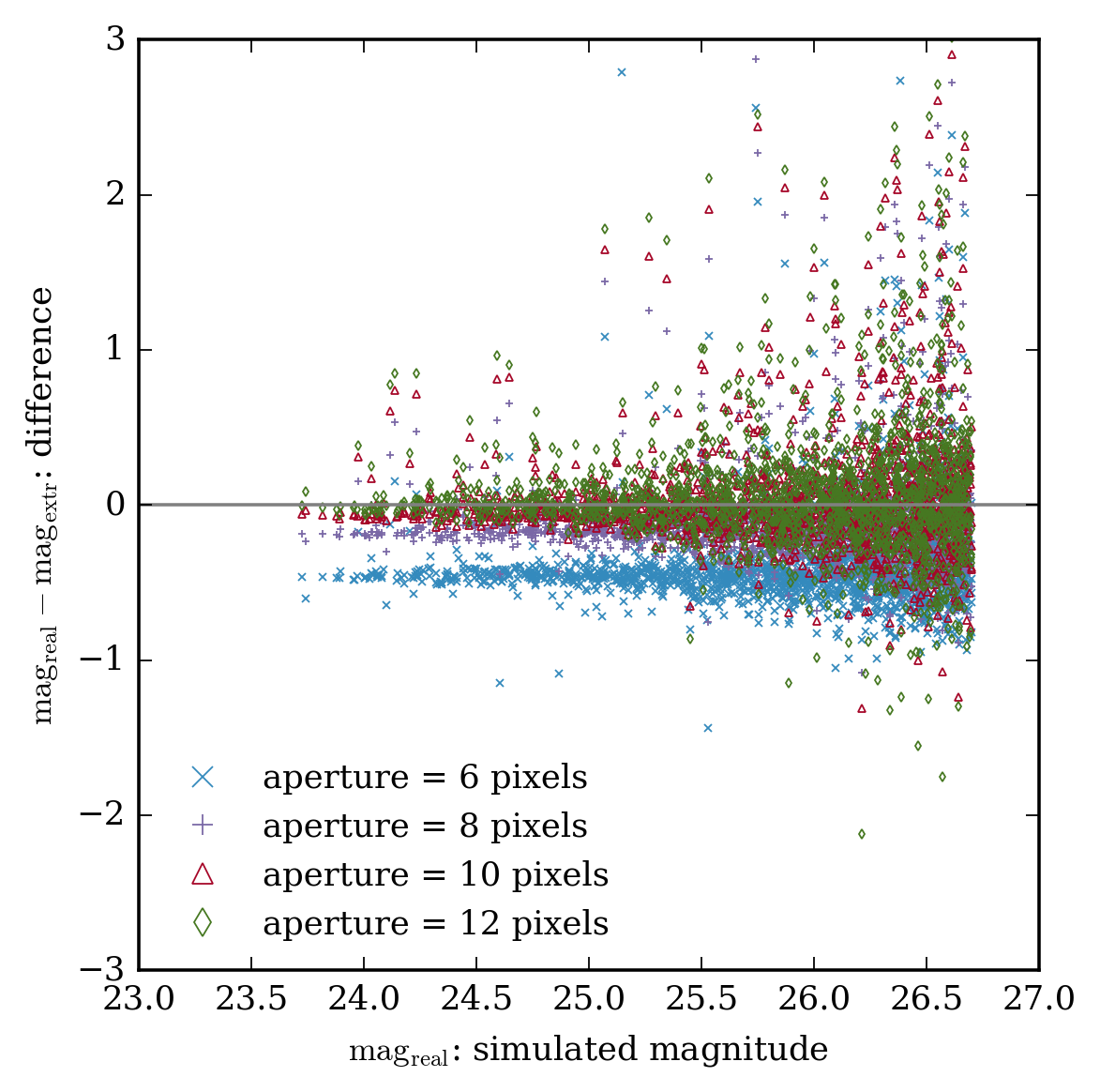}
        \caption{Recovery of the input magnitudes of a simulated emission-line population using different SExtractor aperture sizes.}
        \label{fig:aper_sel}
\end{center}
\end{figure}

We extracted emission-line objects using the object detection algorithm SExtractor \citep{1996A&AS..117..393B} in dual-image mode. Objects were detected if the flux value in at least 20 adjacent pixels is higher than $1.2\times\sigma_{\mathrm{rms}}$ in the on-band image. We chose an aperture radius of 12 pixels, since this is the smallest aperture in which we can recover the magnitudes of a simulated emission-line population (see Fig. \ref{fig:aper_sel}). The detection in dual-image mode was carried out on the on-band image and the extracted aperture magnitudes are denoted with $m_{\mathrm{n}}$. In order to determine the colour of the emission-line objects, the broad-band magnitudes $m_{\mathrm{b}}$ were extracted in the same apertures as on the narrow-band image. The colour is defined as $m_{\mathrm{n}} - m_{\mathrm{b}}$. 

In order to determine the limiting magnitude of our sample, we simulated a synthetic point-like population onto the on-band image using the IRAF task \texttt{mkobjects}. The synthetic population follows a Planetary Nebula Luminosity Function (PNLF) as detailed in \citet[][see also Sect. \ref{sec:PNLF}]{1989ApJ...339...53C} and the objects have a Gaussian PSF profile as detailed in Sect. \ref{ssec:data_red}. The limiting magnitude is defined as the magnitude at which the recovery fraction of the simulated objects drops below $50\%$. The limiting $AB$ magnitude of the on-band image is $m_{\mathrm{n,lim}} = 26.8$ or $m_{\mathrm{5007,lim}} = 29.3$ (see also column 3 of Table \ref{tab:photometric}).

\subsection{Masking of bad or noisy regions}
After identifying emission-line objects on the images, the image regions which were affected by dithering or saturation are masked on the on- and off-band images. Dithering leads to different exposure depths at the image edges. Due to the combination of the 10 Suprime Cam CCDs, the same effect also affects columns at the borders of the individual CCDs. In order to mask regions with a high background value (e.g. due to charge transfer or saturated stars) we used the rms-background map created by SExtractor and create a pixel-mask with all values higher than 1.5 times the median background. We also excluded objects within a major-axis radius of $r_{\mathrm{major}} = 159^{\prime\prime}$, corresponding to a physical distance of $13$ kpc from the galaxy's centre, where the background due to the galaxy's light is higher than the threshold defined above. This reduces the survey area by 4\% to $0.244\;\mathrm{deg}^2$.

\subsection{Colour selection}
\label{ssec:colour_sel}
We selected PN candidates based on their position on the $m_{\mathrm{n}}$ versus $m_{\mathrm{n}} - m_{\mathrm{b}}$ colour-magnitude diagram (CMD). Candidates were objects with a colour excess $m_{\mathrm{n}} - m_{\mathrm{b}} < -1$ that are brighter than the limiting $AB$ magnitude in the narrow band $m_{\mathrm{n,lim}} = 26.8$, which corresponds to $m_{\mathrm{5007,lim}} = 29.3$. The colour excess corresponds to an equivalent width $EW_{\mathrm{obs}} = 110\; \AA$ \citep{2000ApJ...542...18T} and was chosen in order to limit contamination from background galaxies (see discussion in Sect. \ref{ssec:lyalpha}).

To evaluate the contamination by faint stellar sources that fall below the adopted colour excess due to photometric errors, we simulated a continuum population onto the on- and off-band images and calculate the $99\%$ and $99.9\%$ limits based on the extracted CMD of the population. Below these limits, the probability of detecting stars was reduced to the $1\%$ and $0.1\%$ level respectively. Objects with colour-excess above the $99.9\%$ limit are not considered for further analysis. The CMD with the selection criteria overplotted is shown in Fig. \ref{fig:cmdsel}. [OIII] sources for which no broad-band magnitude can be measured are assigned a magnitude of $m_\mathrm{b} = 28.7$, which corresponds to the flux from an [OIII] emission of $m_\mathrm{n} = m_{\mathrm{n,lim}}$ observed through a broad-band $V$ filter \citep{1997MNRAS.284L..11T}. These sources are located on the diagonal line in Fig. \ref{fig:cmdsel}. 

\subsection{Point-like versus extended sources}
\label{app:pointlike}
In order to distinguish point-like sources from extended ones (e.g. background galaxies or other extended objects with strong [OIII] emission), we again first considered the simulation of the PNe on the on-band image and analyse their light distribution. Based on the simulated population, we considered sources as point-like if they satisfied the following two criteria:
these sources (i) have a half-light radius of $1 < R_{\mathrm{h}} < 3.5$ pixel, where the upper limit corresponds to the $95\%$-percentile of the simulated population, and
(ii) they fall in the region where the difference between $m_{\mathrm{n}}$ and $m_{\mathrm{core}}$ is within the $95\%$-limits of the simulated population. The core magnitude $m_{\mathrm{core}}$ was measured in a fixed circular aperture with a radius of four pixels at the same position of the main detection. For point-like objects, the difference between the two magnitudes is expected to be constant as a function of magnitude, for extended sources it varies.
The results of this test are shown in Fig. \ref{fig:pointsource}. The objects which fulfil these criteria along with the ones detailed in the previous section are indicated in the CMD of Fig. \ref{fig:cmdsel}.

\subsection{Catalogue completeness}
\label{app:completeness}
As we selected PN-candidates on flux-ratio-based criteria, we might not have detected PNe that fell below the flux-threshold due to photometric errors. In order to quantify the detection incompleteness, we simulated a PN-population onto the unmasked regions of the on-band image and determined how many sources we retrieved as a function of magnitude. For sources brighter than the limiting magnitude, the average detection completeness was $80\%$. 
The detection completeness as a function of magnitude is displayed in Table \ref{tab:photometric}. We then used the simulated population brighter than the limiting magnitude to determine the colour completeness of our selected sample. This is needed, 
as there will be genuine PNe which lie in CMD regions outside of the colour-selection criteria because of photometric errors. The average colour completeness of our selected sample from 23.5 to 26.8 $AB$ magnitudes was $81\%$ and the magnitude-binned values are presented in Table \ref{tab:photometric}. 

Similarly, we determined the spatial completeness by determining the recovery fraction of the simulated population within the limiting magnitude as a function of major-axis radius (see Table \ref{tab:spatial}).

\begin{table}
\caption{Photometric completeness: colour completeness $c_\mathrm{colour}$ and detection completeness $c_\mathrm{detection}$ for a range of narrow-band magnitudes $m_\mathrm{n}$}
\label{tab:photometric}
\centering
\begin{tabular}{ccc}
\hline\hline
$m_\mathrm{n}$ & $c_\mathrm{colour}$ & $c_\mathrm{detection}$ \\
$[\mathrm{mag}]$ &  &  \\
\hline
23.97 & 0.80 & 0.92 \\
24.23 & 0.87 & 1.00 \\
24.48 & 0.84 & 0.83 \\
24.74 & 0.84 & 0.90 \\
25.00 & 0.86 & 0.82 \\
25.25 & 0.84 & 0.82 \\
25.51 & 0.82 & 0.77 \\
25.77 & 0.81 & 0.77 \\
26.03 & 0.78 & 0.67 \\
26.28 & 0.74 & 0.61 \\
26.54 & 0.70 & 0.65 \\
\hline
\end{tabular}
\end{table}

\begin{table}
\caption{Spatial completeness $c_\mathrm{spatial}$ for a range of major-axis radii $r_\mathrm{major}$. $A$ denotes the un-masked bin area and $a_frac$ the unmasked fraction of a complete elliptical bin.}  
\label{tab:spatial}
\centering
\begin{tabular}{cccc}
\hline\hline
$r_\mathrm{major}$ & $A$ & $c_\mathrm{spatial}$ & $a_\mathrm{frac}$ \\
$[{}^{\circ}]$ & $[{}^{\circ^2}]$ &  &  \\
\hline 
0.08 & 0.0069 & 0.69 & 0.04 \\
0.10 & 0.0084 & 0.79 & 0.94 \\
0.13 & 0.0140 & 0.83 & 0.98 \\
0.15 & 0.0160 & 0.88 & 0.98 \\
0.19 & 0.0260 & 0.93 & 0.99 \\
0.22 & 0.0290 & 0.86 & 0.95 \\
0.28 & 0.0500 & 0.93 & 0.80 \\
0.40 & 0.0790 & 0.88 & 0.44 \\
\hline 
\end{tabular}
\end{table}

\section{The $z = 3.1$ Ly-$\alpha$ luminosity function}
\label{sec:lya}
\citet{2007ApJ...667...79G} carried out a deep survey for $z = 3.1$ Ly-$\alpha$ emission-line galaxies.
The LF is characterised by a Schechter-function \citep{1976ApJ...203..297S} 
        \begin{equation}
                \phi(L/L^{\star}) \,\mathrm{d}(L/L^{\star}) \propto (L/L^{\star})^{\alpha} \exp(-L/L^{\star})\, \mathrm{d}(L/L^{\star})
        ,\end{equation}
with parameters $\log L^{\star} = 42.66\,\mathrm{erg}\,\mathrm{s}^{-1}$, $\alpha = -1.36$ and normalisation $\phi^{\star} = 1.28\times10^{-3}$. The survey has a limiting magnitude of $m_{5007\mathrm{,lim}} = 28.3$, which is somewhat brighter than the limiting magnitude of our study. We extrapolated the LF down to the depth of our survey, which is $m_{5007\mathrm{,lim}} = 28.8$ (cf. Sect. \ref{ssec:vis_insp}). 

The survey area of $0.244\,\mathrm{deg}^2$ corresponds to a comoving volume of $1.84\times 10^5\,\mathrm{Mpc}$. However, as \citetalias{2007ApJ...667...79G} state, the effective survey volume is roughly 25$\%$ smaller when taken through a non-square filter bandpass. Taking this into account, the total number of Ly-$\alpha$ emitters that we expect in our survey area is 310.

The clustering correlation length is $r_0 = 3.6\;\mathrm{Mpc}$, which corresponds to $r_0^{\prime} = 7.9^{\prime}$ at the distance of the Virgo cluster. We can thus expect about $20\%$ variation in the LF due to large-scale cosmic variance \citep{2007ApJ...671..278G, 2004ApJ...600L.171S}. 
Accounting for this variation, $(29.8\pm6.0)\%$ of the objects in the completeness-corrected sample are predicted to be Ly-$\alpha$ emitting background galaxies. Fig. \ref{fig:PNLF_nocorr} shows the \citetalias{2007ApJ...667...79G} Ly-$\alpha$ LF alongside with the LF of the selected PNe.

\end{document}